\documentclass[10pt,letterpaper]{article}
\pdfoutput=1

\usepackage{jcappub}
\usepackage{amsmath,amssymb,color}
\usepackage{enumerate,xstring}
\usepackage{jcappub}
\usepackage{amsmath}
\usepackage{amsfonts}
\usepackage{amssymb}
\usepackage{graphicx}
\usepackage{epsfig}
\usepackage{color}
\usepackage{multirow}
\usepackage{graphicx}
\usepackage{bigints}
\usepackage{todonotes,bm,etoolbox}
\usepackage{calligra}
\usepackage{hyperref}
\usepackage{tabularx}
\usepackage{booktabs}
\usepackage{subfigure}

\usepackage{tikz}

\def\ba#1\ea{\begin{align}#1\end{align}}
\def\bea{\begin{eqnarray}}
\def\eea{\end{eqnarray}}
\def\be{\begin{equation}}
\def\ee{\end{equation}}

\def\({\left(}
\def\){\right)}
\def\[{\left[}
\def\]{\right]}

\def\<{\left\langle}
\def\>{\right\rangle}

\def\comment#1{}

\def\eps{\epsilon}

\renewcommand{\v}[1]{\bm{#1}}

\def\vx{\v{x}}
\def\vk{\v{k}}

\newcommand{\tng}{\textsc{IllustrisTNG}}

\newcommand{\perm}[1]{ \expandafter\ifstrempty\expandafter{#1} {\mbox{perm.}} {\mbox{$#1$ perm.}} }

\def\O{\mathcal{O}}

\def\P{\mathcal{P}}

\newcommand{\tngbig}{L_{\rm box}\approx800{\rm Mpc}}

\newcommand{\refeq}[1]{Eq.~(\ref{eq:#1})} 

\definecolor{RedWine}{rgb}{0.743,0,0}
\definecolor{RoyalBlue}{rgb}{0.25,.41,.88}
\definecolor{ForestGreen}{rgb}{.13,.54,.13}
\definecolor{Goldenrod}{rgb}{.85,.65,.13}

\newcommand{\bq}{\begin{eqnarray}}
\newcommand{\eq}{\end{eqnarray}}

\title{Galaxy bias from forward models: \huge linear and second-order bias of IllustrisTNG galaxies}

\author[a,b]{Alexandre Barreira}
\author[c,d,e]{Titouan Lazeyras}
\author[f]{Fabian Schmidt}

\affiliation[a]{\small Excellence Cluster ORIGINS, Boltzmannstra\ss e 2, 85748 Garching, Germany}
\affiliation[b]{\small Ludwig-Maximilians-Universit\"at, Schellingstra\ss e 4, 80799 M\"unchen, Germany}
\affiliation[c]{\small SISSA, Via Bonomea 265, 34136 Trieste, Italy}
\affiliation[d]{\small INFN, Sezione di Trieste, Via Bonomea 265, 34136 Trieste, Italy }
\affiliation[e]{\small IFPU, Institute for Fundamental Physics of the Universe, via Beirut 2, 34151, Trieste, Italy}
\affiliation[f]{\small Max-Planck-Institut f\"ur Astrophysik, Karl-Schwarzschild-Stra\ss e~1, 85748 Garching, Germany}

\emailAdd{alex.barreira@origins-cluster.de}
\emailAdd{tlazeyra@sissa.it}
\emailAdd{fabians@mpa-garching.mpg.de}

\date{\today}

\abstract{We use field-level forward models of galaxy clustering and the EFT likelihood formalism to study, for the first time for self-consistently simulated galaxies, the relations between the linear $b_1$ and second-order bias parameters $b_2$ and $b_{K^2}$. The forward models utilize all of the information available in the galaxy distribution up to a given order in perturbation theory, which allows us to infer these bias parameters with high signal-to-noise, even from relatively small volumes ($L_{\rm box} = 205{\rm Mpc}/h$). We consider galaxies from the \tng\ simulations, and our main result is that the $b_2(b_1)$ and $b_{K^2}(b_1)$ relations obtained from gravity-only simulations for total mass selected objects are broadly preserved for simulated galaxies selected by stellar mass, star formation rate, color and black hole accretion rate. We also find good agreement between the bias relations of the simulated galaxies and a number of recent estimates for observed galaxy samples. The consistency under different galaxy selection criteria suggests that theoretical priors on these bias relations may be used to improve cosmological constraints based on observed galaxy samples. We do identify some small differences between the bias relations in the hydrodynamical and gravity-only simulations, which we show can be linked to the environmental dependence of the relation between galaxy properties and mass.  We also show that the EFT likelihood recovers the value of $\sigma_8$ to percent-level from various galaxy samples (including splits by color and star formation rate) and after marginalizing over 8 bias parameters. This demonstration using simulated galaxies adds to previous works based on halos as tracers, and strengthens further the potential of forward models to infer cosmology from galaxy data.}

\begin{document}

\maketitle

\section{Introduction}
\label{sec:intro}

The relation between the distribution of galaxies and of the underlying distribution of mass and energy in the Universe is known as {\it galaxy bias}, and it is a central ingredient in cosmological inference analyses using galaxy clustering (see Ref.~\cite{biasreview} for a comprehensive review). The deterministic part of this relation can be written as 
\bq\label{eq:biasexp_intro}
n_g(\vx, z) = \bar{n}_g(z) \left[1 + \sum_{\O}b_\O(z)\O(\vx, z)\right],
\eq
where $n_g(\vx, z)$ is the rest-frame number density of galaxies at position $\vx$ and redshift $z$,  and $\bar{n}_g(z)$ is its cosmic average. The sum runs over all long-wavelength perturbations $\O$ of mass and energy in the Universe that can influence galaxy formation. Each of these is multiplied by a {\it bias parameter} $b_\O$ that physically describes the {\it response} of the local number density of galaxies to changes in the amplitude of the perturbations $\O$. From an effective field theory (EFT) perspective \cite{baumann, 2012JHEP...09..082C}, the perturbations $\O$ are regarded as sufficiently long-wavelength to be able to be described by perturbation theory \cite{Bernardeau/etal:2002}, with the bias parameters effectively absorbing all of the complicated details that govern galaxy formation and evolution. The scatter around the relation \refeq{biasexp_intro} due to small-scale perturbations is taken into account by another set of free parameters called {\it stochastic parameters}.  

Galaxy bias is important to study for at least two reasons. First, it can lead to insights about galaxy formation and evolution, and in particular its dependence on the long-wavelength environment, if observational determinations of the bias parameters can be compared with predictions from different models of galaxy formation. Second, this also opens the possibility to design theoretical priors for the bias parameters and relations between them, which can then be used to reduce the volume of the parameter space explored in cosmological constraint analyses using galaxy clustering data \cite{2016arXiv160200674B, 2020PhRvD.102l3521W}. In these analyses, the galaxy bias parameters often display degeneracies with the cosmological parameters, and so the tighter our prior knowledge on galaxy bias, the tighter our bounds on the cosmological parameters.

The most popular and best-studied bias parameters are those associated with perturbations that are powers of the matter density contrast field $\delta_m(\vx, z)$, which enter the bias expansion as $b_1\delta_m$, $(b_2/2)\delta_m^2$, $(b_3/3!)\delta_m^3$, etc.~\cite{fry/gaztanaga:1983, 2010ApJ...724..878T, lazeyras/etal, 2017MNRAS.465.2225H}; these bias parameters are called local-in-matter-density (LIMD) parameters. It is well known also that galaxies respond to large-scale tidal fields $K_{ij}(\vx, z)$, with the leading-order contribution to Eq.~(\ref{eq:biasexp_intro}) being $b_{K^2} K^2$ \cite{mcdonald/roy:2009, sheth/chan/scoccimarro:2012, chan/scoccimarro/sheth:2012, baldauf/etal:2012, MSZ, 2018JCAP...09..008L, saito/etal:14, 2018JCAP...07..029A, 2020PhRvD.102j3530E, 2021arXiv210206902E} (see Eq.~(\ref{eq:operators}) below). Other types of perturbations involve higher-than-second-order derivatives of the gravitational potential, for example $\O = \nabla^2\delta_m(\vx, z)$ \cite{2019JCAP...11..041L, 2020PhRvD.102j3530E, 2021arXiv210206902E}; primordial gravitational potential perturbations $\O  = \phi(\vx)$ in case of primordial non-Gaussianity \cite{dalal/etal:2008, giannantonio/porciani:2010, assassi/baumann/schmidt, 2020JCAP...12..013B, 2020arXiv201014523M, 2020JCAP...12..031B}; relative baryon-CDM density and velocity perturbations \cite{tseliakhovich/hirata:2010, 2016PhRvD..94f3508S, 2020JCAP...02..005B,  2021JCAP...03..023K}; and perturbations of the ionizing radiation field in the Universe \cite{2017PhRvD..96h3533S, 2019JCAP...05..031C}. The vast majority of bias studies are done using gravity-only simulations and focus on the bias of dark matter halos. While these have led to a robust knowledge of halo bias, including the establishment of precise relations between different bias parameters, it is important to check the extent to which these relations also work for the case of self-consistently simulated galaxies, whose formation depends on more physics other than gravity. For example, in the context of the bias parameters relevant for constraints of primordial non-Gaussianity and compensated baryon-CDM isocurvature perturbations, Refs.~\cite{2020JCAP...02..005B, 2020JCAP...12..013B} showed recently that there are important differences between the bias values of galaxies selected by stellar mass, compared to what one would naively expect based on the bias of halos selected by their total mass; Refs.~\cite{2020JCAP...07..049B, 2020JCAP...12..031B} subsequently showed that these differences can have a dramatic impact on the resulting cosmological constraints. 

The study of galaxy bias using self-consistent galaxy formation simulations has only been made possible recently with the advent of efficient numerical codes to simulate galaxy formation and evolution in cosmologically representative volumes; examples include the {\sc Illustris} \cite{2014MNRAS.444.1518V}, {\sc EAGLE} \cite{2015MNRAS.446..521S, 2017arXiv170609899T}, {\sc Magneticum} \cite{2014MNRAS.442.2304H}, {\sc BAHAMAS} \cite{2017MNRAS.465.2936M}, {\sc Horizon-AGN} \cite{2014MNRAS.444.1453D} and {\sc IllustrisTNG} \cite{Pillepich:2017jle, 2017MNRAS.465.3291W, Nelson:2018uso} projects. The philosophy of these projects consists of coupling the action of gravity and hydrodynamics with physical processes like gas cooling, star formation, and stellar and black hole feedback, which are implemented as coarse-grained effective models that are calibrated to reproduce a number of key observables like the stellar mass function at low-$z$, galaxy sizes, gas fractions, etc.~(different projects adopt different calibration data sets). In this paper we will use the \tng\ simulation data to study the first two LIMD parameters $b_1$, $b_2$ and the leading-order tidal bias parameter $b_{K^2}$. These parameters contribute to the galaxy power spectrum at the 1-loop level, as well as the galaxy bispectrum at tree-level, and so they are critical in cosmological constraint analyses using these two galaxy statistics. Using gravity-only separate universe simulations, Ref.~\cite{lazeyras/etal} found that dark matter halos follow a relation $b_2(b_1)$ with only small scatter, at different redshifts and mass scales. The relation between $b_{K^2}$ and $b_1$ is currently not as precisely established, even for dark matter halos, although a number of recent studies \cite{saito/etal:14, 2018JCAP...09..008L, 2018JCAP...07..029A, 2020PhRvD.102j3530E, 2021arXiv210206902E} have been progressively suggesting that a $z$-independent relation might exist as well, at least when the objects are selected according to their total mass. Our main goal here using the \tng\ simulations is to investigate, for the first time, the form of these relations for self-consistently simulated galaxies, and as a function of different selection criteria including total mass, stellar mass, star formation rate, galaxy color and central black hole accretion rate.

Another novel aspect of our analyses here is the application, for the first time, of field-level forward models of galaxy clustering to study these galaxy bias parameters in simulations. The central physical ingredients in this approach are (i) a forward model to evolve the initial matter distribution to the final state; (ii) a galaxy bias model (\refeq{biasexp_intro}) to construct the galaxy distribution out of the final matter distribution; and (iii) a likelihood function to compare the forward-evolved galaxy field with the simulated one. Forward models are still in their infancy, but rapid progress is being made \cite{2010MNRAS.409..355J, 2013MNRAS.432..894J, 2014ApJ...794...94W, 2015MNRAS.446.4250A, 2016ApJ...831..164W, 2017JCAP...12..009S, 2019PhRvD.100d3514S, 2019JCAP...11..023M, 2020arXiv201203334S, 2019A&A...625A..64J, 2019arXiv190906396L, 2019JCAP...01..042S, 2020JCAP...01..029E, 2020JCAP...11..008S, 2020arXiv200914176S, 2021JCAP...03..058N}. A key advantage of forward models is that the inference is made directly at the level of the {\it actually observed} galaxy density field, which maximizes the extraction of the information available up to a given order in perturbation theory, in contrast to relying for example on a handful of summary statistics like $N$-point functions. Applications of forward models to real data are very numerically intensive as, in addition to the bias parameters, one must sample also over the cosmological parameters and phases of the initial matter distribution (if the latter is defined on a grid with $100$ nodes, this amounts already to $100^3$ parameters). In our case here, crucially since our {\it observed} galaxy sample is in fact a simulated one, we already know the initial conditions and cosmology used to run the simulation, which cancels sample variance to a large degree and leaves us only with the much more reduced galaxy bias parameter space to fit for.

In this paper we build specifically on the recent developments made in Refs.~\cite{2019JCAP...01..042S, 2020JCAP...01..029E, 2020JCAP...11..008S, 2020arXiv200914176S} on forward models, including the derivation of the crucial likelihood function using the EFT formalism \cite{2019JCAP...01..042S, 2020JCAP...04..042C, 2020JCAP...07..051C}. This paper is accompanied by another one using the same methodology, but focused on the assembly bias signal of $b_2$ and $b_{K^2}$ of dark matter halos \cite{Lazeyras:2021dar} (i.e.~the dependence of these bias parameters on halo properties beyond their total mass). The rest of this paper is organized as follows. In Sec.~\ref{sec:method} we outline the basics of the forward modeling and EFT likelihood methodology, and describe also the \tng\ simulation data that we use to study $b_2$ and $b_{K^2}$. In Sec.~\ref{sec:biasresults} we present and discuss our main numerical results on the bias parameters. Section~\ref{sec:s8results} shows an application of the forward modeling approach to infer the cosmological parameter $\sigma_8$ using simulated galaxies as tracers. Finally, we summarize and conclude in Sec.~\ref{sec:summary}.

\section{Methodology}
\label{sec:method}

In this section, we describe the forward-modeling and EFT likelihood methodology that we use in this paper. This methodology has been developed and tested in a series of recent papers \cite{2019JCAP...01..042S, 2020JCAP...01..029E, 2020JCAP...04..042C, 2020JCAP...07..051C, 2020JCAP...11..008S, 2020arXiv200914176S, 2020arXiv201209837S}, and so we limit ourselves to simply briefly outlining its main aspects; the various numerical setup choices described below are also largely inspired by the tests performed in these past works (see also Ref.~\cite{2021JCAP...03..058N} for a study of the relative importance of the different physical ingredients that enter inference analyses with forward models). We also describe the galaxy simulation data we use from the \tng\ simulations to obtain our results.

\subsection{Forward models and the EFT likelihood}
\label{sec:fwdmodel}

The goal of forward models of galaxy clustering is to sample the \emph{likelihood} $\P\big(\delta_g | \{\theta\}, \{b_\O\}, \delta_{m,\rm in}\big)$ that describes the probability of observing a galaxy density contrast field $\delta_g$, given a set of cosmological parameters $\{\theta\}$, a set of bias parameters $\{b_\O\}$ and a realization $\delta_{m,\rm in}$ of the initial matter density contrast out of which the observed galaxy sample formed. Schematically, this works as follows:
\begin{enumerate}[i]

\item Evolve $\delta_{m,\rm in}$ under gravity in a given cosmology to generate the foward-evolved final matter distribution, $\delta_{m,\rm fwd}[\{\theta\}, \delta_{m,\rm in}]$. Here, we take third-order Lagrangian perturbation theory (3LPT) as the \emph{gravity model} \cite{2020arXiv201209837S}.

\item Construct a realization of the forward-evolved galaxy distribution out of the final matter distribution using a deterministic galaxy bias expansion, $\delta_{g, \rm det}[\delta_{m, \rm fwd}, \{b_\O\}]$.

\item Iterate over the initial conditions $\delta_{m, \rm in}$, the cosmological parameters $\{\theta\}$ and the bias parameters $\{b_\O\}$, to sample the likelihood $\P\big(\delta_g | \delta_{g, {\rm det}}\big) \equiv \P\big(\delta_g | \{\theta\}, \{b_\O\}, \delta_{m,\rm in}\big)$.

\end{enumerate}

In this paper, we are interested in studying the bias parameters of \textit{simulated} galaxies, which lets us fix the cosmological parameters $\{\theta\}$ and initial conditions field $\delta_{m, \rm in}$ to those used to run the simulation, and thus fit very efficiently for the $\{b_\O\}$ parameter space. We work with the following likelihood function in Fourier space (we distinguish Fourier- from configuration-space quantities by their arguments), which has been derived using EFT in Refs.~\cite{2019JCAP...01..042S, 2020JCAP...01..029E, 2020JCAP...04..042C, 2020JCAP...11..008S, 2020JCAP...07..051C} (see also Ref.~\cite{2020arXiv200914176S} for its version in configuration space)
\bq\label{eq:eftlike}
-2{\rm ln}\P\big(\delta_g | \delta_{g, {\rm det}}\big) = \int_{|\vk| < \Lambda} \frac{{\rm d}^3\vk}{(2\pi)^3} \Bigg[\frac{\big|\delta_{g}(\vk) - \delta_{g, {\rm det}}(\vk)\big|^2}{P_\eps(k)} + {\rm ln}\big(2\pi P_\eps(k)\big)\Bigg].
\eq
In this equation, the integral is performed only up to a maximum cutoff wavenumber $\Lambda$. The application of this cutoff ensures that only perturbative modes enter the likelihood evaluation, in keeping with the EFT approach to galaxy clustering and bias expansion. Below we will investigate the impact of different choices for the cutoff $\Lambda$, but which we note must always be lower than the nonlinear scale $k_{\rm NL}$ at a given redshift ($k_{\rm NL} \approx 0.3 h/{\rm Mpc}$ at $z=0$). Importantly, to regularize loop integrals that would otherwise involve smaller-scale, non-perturbative modes, this cutoff must also be applied to the initial conditions field $\delta_{m, \rm in}$ before evolving it to the final time \cite{2020JCAP...11..008S}.

The variance $P_\eps(k)$ in Eq.~(\ref{eq:eftlike}) is what takes into account the stochasticity of galaxy formation, and it is in general a function of wavenumber \cite{2020JCAP...04..042C, 2020JCAP...07..051C} (see also Ref.~\cite{hamaus/etal:2010} for a study of halo stochasticity using a similar likelihood function). Here, we will follow the same strategy as in Ref.~\cite{2020JCAP...11..008S} and consider only the leading order, constant contribution $P_\eps(k) \approx P_\eps^{\{0\}}$, where $P_\eps^{\{0\}}$ is a parameter that is also sampled. This approximation is expected to have a negligible impact in our numerical results \cite{2020JCAP...11..008S, 2020arXiv200914176S}. Finally, in practice, the integral in Eq.~(\ref{eq:eftlike}) is replaced by a sum over the nodes of a regular, cubic grid covering the simulation volume onto which all fields are discretized (all grids we use here have $N_{\rm grid} = 512$ nodes on a side).

\subsection{The galaxy bias expansion}
\label{sec:biasexp}

Given the forward-evolved matter density contrast $\delta_{m,\rm fwd}(\vx)$, we construct the corresponding forward-evolved galaxy density field using the deterministic galaxy bias expansion
\bq\label{eq:biasexp_E}
\delta_{g, {\rm det}}(\vx) = \sum_\O b_\O \O_\Lambda(\vx),
\eq
where the subscript $_\Lambda$ in $\O_\Lambda(\vx) \equiv \O[\delta_{m, \rm fwd, \Lambda}(\vx)]$ indicates these are terms constructed out of $\delta_{m, \rm fwd, \Lambda}(\vx)$, which is the sharp-$k$, low-pass filtered version of $\delta_{m,\rm fwd}(\vx)$. Here, we work at third-order in the bias expansion and consider the following set of 8 operators (see Sec.~2 of Ref.~\cite{biasreview} for a comprehensive derivation)
\bq\label{eq:operators}
\O \in \big\{\delta_m, \delta_m^2, K^2, \delta_m^3, \delta_m K^2, K^3, O_{\rm td}, \nabla^2\delta_m\big\},
\eq
where $K_{ij} = \big(\partial_i\partial_j/\nabla^2 - \delta_{ij}/3\big)\delta_m$, $K^2 = K_{ij}K^{ij}$, $K^3 = K_{ij}K^{jk}K_k^i$ and $O_{\rm td} = (8/21)K_{ij}(\partial_i\partial_j/\nabla^2)\big(\delta_m^2-(3/2)K^2\big)$. The first 7 of these operators are all that exist up to third order involving terms with second-order derivatives of the gravitational potential (i.e., density and tidal fields). In addition, we consider also one higher-derivative term $\nabla^2\delta$, which according to the strategy described in Ref.~\cite{2020arXiv200914176S}, is the only such operator that is relevant at the order we work with here. More specifically, the relevance of the higher-derivative operators is controlled by the nonlinear scale $k_{\rm NL}$, the cutoff $\Lambda$ and the spatial nonlocality scale of the tracers $R_*$. Below, we will show results for different redshifts (and consequently different $k_{\rm NL}$) and values of $\Lambda$, which impacts the ranking of the higher-derivative terms. For simplicity, and in order to ensure we always use the same number of operators when we vary $\Lambda$ and $z$ below, we evaluate the importance of the higher-derivative terms at the following fixed values: $\Lambda = 0.14 h/{\rm Mpc}$, $k_{\rm NL}(z \approx 0) = 0.25 h/{\rm Mpc}$ and $R_* = 5{\rm Mpc}/h$ (which is roughly the Lagrangian radius of the tracers we consider). These reasonable values imply that $\nabla^2\delta_m$ is the only relevant higher-derivative operator \cite{2020arXiv200914176S}.

In practice, the construction of $\delta_{g, {\rm det}}(\vx)$ takes place as follows. The initial conditions field $\delta_{m, \rm in}(\vx)$ (which in our case are the initial conditions at $z=127$ used to run the \tng\ simulations) is discretized on a grid and filtered out of modes with $k > \Lambda$ (this is done by going back-and-forth between configuration and Fourier space). The filtered field is evolved to the desired final redshift using third-order LPT to generate the final mass distribution $\delta_{m, \rm fwd}(\vx)$ (Ref.~\cite{2020arXiv201209837S} showed that the power spectrum of the 3LPT field is accurate to well within $1\%$ for all scales used in this paper; note that we never identify structures in the forward-evolved field). The latter, again filtered with a sharp-$k$ filter on the scale $\Lambda$, is subsequently used to construct the operators $\O$ in Eq.~(\ref{eq:operators}), which are then added up according to Eq.~(\ref{eq:biasexp_E}) to generate $\delta_{g, {\rm det}}(\vx)$. This field, together with the {\it observed} galaxy sample $\delta_g(\vx)$ (which in our case are \tng\ galaxy samples), are then finally transformed to Fourier space to be used in the EFT likelihood of Eq.~(\ref{eq:eftlike}).

We note that the quadratic operators $\delta_m^2, K^2$ are renormalized with respect to $\delta_m$ by subtracting their overlap with $\delta_m$ as described in Ref.~\cite{2019JCAP...01..042S}. This is important in order to be able to interpret the corresponding bias parameters as those that appear in the large-scale $N$-point functions \cite{2019JCAP...01..042S}. We do not perform renormalizations with respect to the quadratic operators themselves, which as we mention below could partly be responsible for small systematic shifts in the inferred bias parameters for higher values of the cutoff $\Lambda$. Our companion paper \cite{Lazeyras:2021dar} discusses the $\Lambda$-dependence of the inferred bias parameters in more detail.

\subsection{The fitting procedure}
\label{sec:fitting}

As we mentioned before, when we use the forward model and the EFT likelihood to fit for the bias parameters we keep the initial conditions, as well as all cosmological parameters fixed to the \tng\ values. To fit for a galaxy bias parameter $b_\O$, we first marginalize over all others in Eq.~(\ref{eq:eftlike}) (which can be done analytically as shown in \cite{2020JCAP...01..029E,2020JCAP...11..008S}), and then search for the maximum of the likelihood in the remainder of the parameter space, which is $\{b_\O, P_\eps^{\{0\}}\}$ (this is done using the {\sc minuit} routines from the {\sc root}\footnote{https://root.cern.ch/} package). The maximum-likelihood value of $b_\O$ is what we quote as the inferred bias parameters in this paper, and our error bars are given by the inverse square root of the curvature of $-2{\rm ln}\P$ in the $b_\O$ direction at the best-fitting points. Note that we do not marginalize over $P_\eps^{\{0\}}$, but this should not have a critical impact in our quoted errors as $P_\eps^{\{0\}}$ and the $b_\O$ are only very weakly correlated. We defer the development of a proper sampling of the full shape of the likelihood for future work, but note that although the minimization procedure is less stable and may sometimes converge to a different local (rather than the global) minimum, we have found this to occur only very rarely in cross-checks we performed to reproduce some results from past works for halos in gravity-only simulations (see Sec.~\ref{sec:lambda} below). In our results throughout, we focus also only on the bias parameters associated with the first three operators in Eq.~(\ref{eq:operators}), $b_1$, $b_2$ and $b_{K^2}$, respectively; note that $b_2$ is related to the coefficient of the second operator in Eq.~(\ref{eq:operators}) as $b_{\delta^2} = b_2/2$. The code naturally returns also estimates of the remaining, higher-order bias parameters, but which we have found were too noisy given the statistical power we could attain with the \tng\ simulation box.

Finally, although in this paper we are mostly focused on the galaxy bias parameters, in Sec.~\ref{sec:s8results} we shall also briefly present inference analyses of the cosmological parameter $\sigma_8$ (the root-mean square amplitude of the density fluctuations on $8{\rm Mpc}/h$ scales). This will be similar to the analyses already presented in Refs.~\cite{2019JCAP...01..042S, 2020JCAP...01..029E, 2020JCAP...11..008S, 2020arXiv200914176S} for dark-matter halos in gravity-only simulations as tracers, except that here we will use galaxies from hydrodynamical simulations. In this part of our analysis, we marginalize analytically over all 8 bias parameters and fit for the parameter $\sigma_8$ by finding the minimum of Eq.~(\ref{eq:eftlike}) at the following fixed values of
\bq\label{eq:sigma8vals}
\frac{\sigma_8}{\sigma_8^{\rm Fiducial}} \in \Big\{0.8,\ 0.9,\ 0.95,\ 0.975,\ 1,\ 1.025,\ 1.05,\ 1.1,\ 1.2\Big\},
\eq
where $\sigma_8^{\rm Fiducial}$ is the value used to run the simulations (note that in this case the only remaining free parameter is $P_\eps^{\{0\}}\}$). The maximum-likelihood value of $\sigma_8$ is then obtained by fitting a parabola to the best-fitting values of $-2{\rm ln}\P$; the estimated error is again given by the inverse square root of the curvature at the minimum. In practice, in the forward model, the different values of $\sigma_8$ are implemented by rescaling the initial density field as $\delta_{m, \rm in} \to \left({\sigma_8}/{\sigma_8^{\rm Fiducial}}\right) \delta_{m, \rm in}$ before evolving it to the final time.

\subsection{Numerical galaxy data from IllustrisTNG}
\label{sec:sims}

We apply the forward modeling methodology described above to galaxy samples from the \tng\ simulations\footnote{https://www.tng-project.org/}, i.e., we use these galaxy catalogues to obtain the $\delta_g(\vk)$ field that enters Eq.~(\ref{eq:eftlike}). The \tng\ model \citep{2017MNRAS.465.3291W, Pillepich:2017jle,Nelson:2018uso} is an effective model of galaxy formation in cosmological simulations that includes prescriptions for gas cooling, star formation and stellar feedback, black hole growth and feedback, and gas (re)ionization. This model is an improved version of its predecessor \textsc{Illustris}, and it broadly matches a series of key galaxy observations including the cosmic star formation rate history, stellar mass function, galaxy sizes, and the gas fractions in galaxies and groups. The interested reader is referred to Refs.~\cite{2018MNRAS.480.5113M, Pillepich:2017fcc, 2018MNRAS.477.1206N, 2018MNRAS.475..676S, Nelson:2017cxy, 2019MNRAS.490.3234N, 2019MNRAS.490.3196P} for the first results with IllustrisTNG.

The \tng\ simulations were run with the moving-mesh hydrodynamic + gravity $N$-body code {\sc AREPO} \citep{2010MNRAS.401..791S, 2016MNRAS.455.1134P}. Here, we consider the publicly-available data \cite{Nelson:2018uso} from the simulations labeled as TNG300-1, which correspond to a box size $L_{\rm box} = 205{\rm Mpc}/h \approx 300{\rm Mpc}$ with $N_p = 2 \times 2500^3$ dark matter tracer particles and initial gas mass elements. In our results below we will consider both the full physics run (which we label as Hydro), as well as its gravity-only counter part (labeled as Gravity). In addition to this box, to perform a few sanity checks of our results, we will consider also a simulation with size $L_{\rm box} = 560{\rm Mpc}/h \approx 800{\rm Mpc}$ and $N_p = 1250^3$ dark matter tracer particles. This simulation, which we label as $\tngbig$, was run also with {\sc AREPO} but for gravity-only (note that this simulation is not part of the public \tng\ release). The cosmological parameters of these simulations are: mean matter density today $\Omega_{m0} = 0.3089$, mean baryon density today $\Omega_{b0} = 0.0486$, mean dark energy density today $\Omega_{\Lambda0} = 0.6911$, dimensionless Hubble rate $h = 0.6774$, primordial scalar spectral index $n_s = 0.967$, and $\sigma_8(z=0) = 0.8159$.

The halos in these simulations have been identified with a friends-of-friends (FoF) algorithm run on the dark matter particle distribution with a linking length $0.2$ times the mean interparticle distance. The subhalos are the substructures found by the {\sc Subfind} \cite{2001MNRAS.328..726S} algorithm inside each FoF halo. The subhalo population includes both the main (central) subhalo, as well as the secondary (satellite) subhalos that orbit around it; following the standard \tng\ nomenclature, we refer to the subhalos with non-zero mass in stars as galaxies. In our results, when quoting the values of the properties of these objects, we always consider the contribution from all their member particles. For example, the stellar mass of a galaxy/halo is the summed mass of all of the stars associated with the corresponding subhalo/FoF halo. Below, we will show results for objects selected by their total mass $M_{\rm t}$, stellar mass $M_*$, $g-r$ dust-uncorrected color \cite{Nelson:2017cxy}, black hole accretion rate $\dot{M}_{\rm BH}$ and specific star formation rate ${\rm sSFR}$ (star formation rate per unit stellar mass). To guarantee we use objects in the simulations that are sufficiently well resolved, we consider only objects that contain at least $100$ star particles.

We will always consider halos and galaxies in the simulations at their real-space positions, i.e.~without taking redshift space distortions into account, which is sufficient to our purpose to focus on the bias parameters of the \tng\ galaxies. In applications to real galaxy samples one would need to go beyond this approximation and use a likelihood function with redshift space distortions information \cite{2021JCAP...01..067C}. Finally, as is standard in analyses of galaxy formation simulations, our conclusions below apply strictly to the specific galaxy physics implementation of the IllustrisTNG model; the dependence (or lack thereof) of our findinds on the details of galaxy formation is an interesting topic to investigate, which we plan to explore in future work.

\section{Results on galaxy bias inference}
\label{sec:biasresults}

In this section we show and discuss our main numerical results on the galaxy bias parameters of the \tng\ galaxies, including the impact of the cutoff value $\Lambda$ on the inferred $b_2(b_1)$ and $b_{K^2}(b_1)$ relations, as well as the impact of different selection criteria for the simulated galaxies. We also discuss and provide a physical interpretation of our results with the aid of the halo model and halo occupation distribution formalisms, and compare the bias relations we obtain for the \tng\ galaxies with recent estimates obtained for observed galaxy samples.

Next, we will show results for objects selected in bins of several properties, including total mass, stellar mass, color,  star formation rate and black hole accretion rate. Bins near the extremes of the distributions of these properties contain fewer objects and the result becomes noisier. We decide which sample results are robust based on the following criterion. For all binned samples, we evaluate the phase-correlation coefficient between galaxies and matter, $r(k) = P_{gm}^{\rm ini}(k)/\sqrt{P_{mm}^{\rm ini}(k)P_{gg}(k)}$, where $P_{mm}^{\rm in}$, $P_{gg}$ and $P_{gm}^{\rm in}$ are the initial matter, galaxy and corresponding cross power spectra, respectively. At low $k$, $r(k)$ is close to unity since the matter power spectrum is much larger than the shot noise, reflecting the strong correlation between the galaxy and initial matter fields on large scales. Since $P^{\rm in}_{mm}, P^{\rm in}_{gm}$ shrink towards higher-$k$ while the shot noise remains constant, the correlation coefficient drops. We thus choose to keep results only for the galaxy samples that satisfy $r(k) > 0.5$ for all $k < \Lambda$. This choice roughly ensures that the samples are still not stochastic (or shot-noise) dominated at the scales of the cutoff $\Lambda$, and thus, that the EFT likelihood formalism can utilize the correlation between the galaxies and matter to infer the bias parameters. 

\subsection{Convergence tests on the cutoff $\Lambda$}
\label{sec:lambda}

\begin{figure}
\centering
\includegraphics[width=\textwidth]{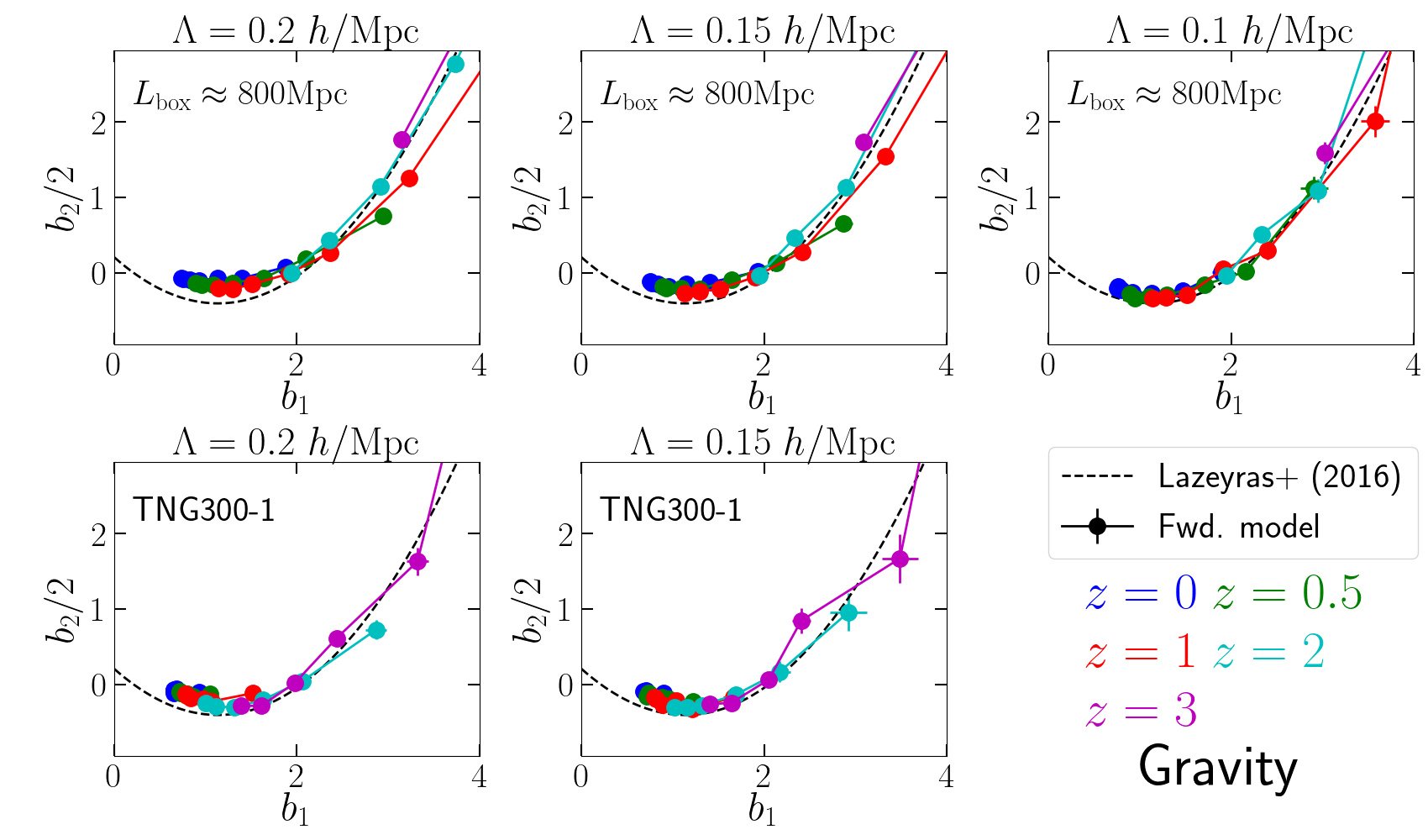}
\caption{The $b_2(b_1)$ relation of dark matter halos in the Gravity runs. The points show the result obtained with the forward model and the EFT likelihood of Eq.~(\ref{eq:eftlike}), with the colors indicating the redshift. This result is for halos selected by their total mass, i.e., each point shows the $b_2$ and $b_1$ values of the halos found in the same total mass bin (the bins can be read from Fig.~\ref{fig:fig_biaspart_Lambda_convergence_0p2_b1_both_tng300_1_tng800_1_group_dmo_totmass}). The different panels are for the two simulation boxes TNG300-1(bottom) and $\tngbig$ (top), as well as different values of $\Lambda$, as labeled. The $\Lambda = 0.1h/{\rm Mpc}$ result for the small box TNG300-1 is too noisy due to fewer available modes and is therefore not shown. The bigger box $\tngbig$ contains also more massive objects, but we focus only on the range of $b_1$ values covered by the smaller TNG300-1 box to facilitate the comparison. The black dashed line shows the fitting formula obtained by Ref.~\cite{lazeyras/etal} using separate universe simulations.}
\label{fig:fig_biaspart_Lambda_convergence_b2_both_tng300_1_tng800_1_group_dmo_totmass}
\end{figure}

\begin{figure}
\centering
\includegraphics[width=\textwidth]{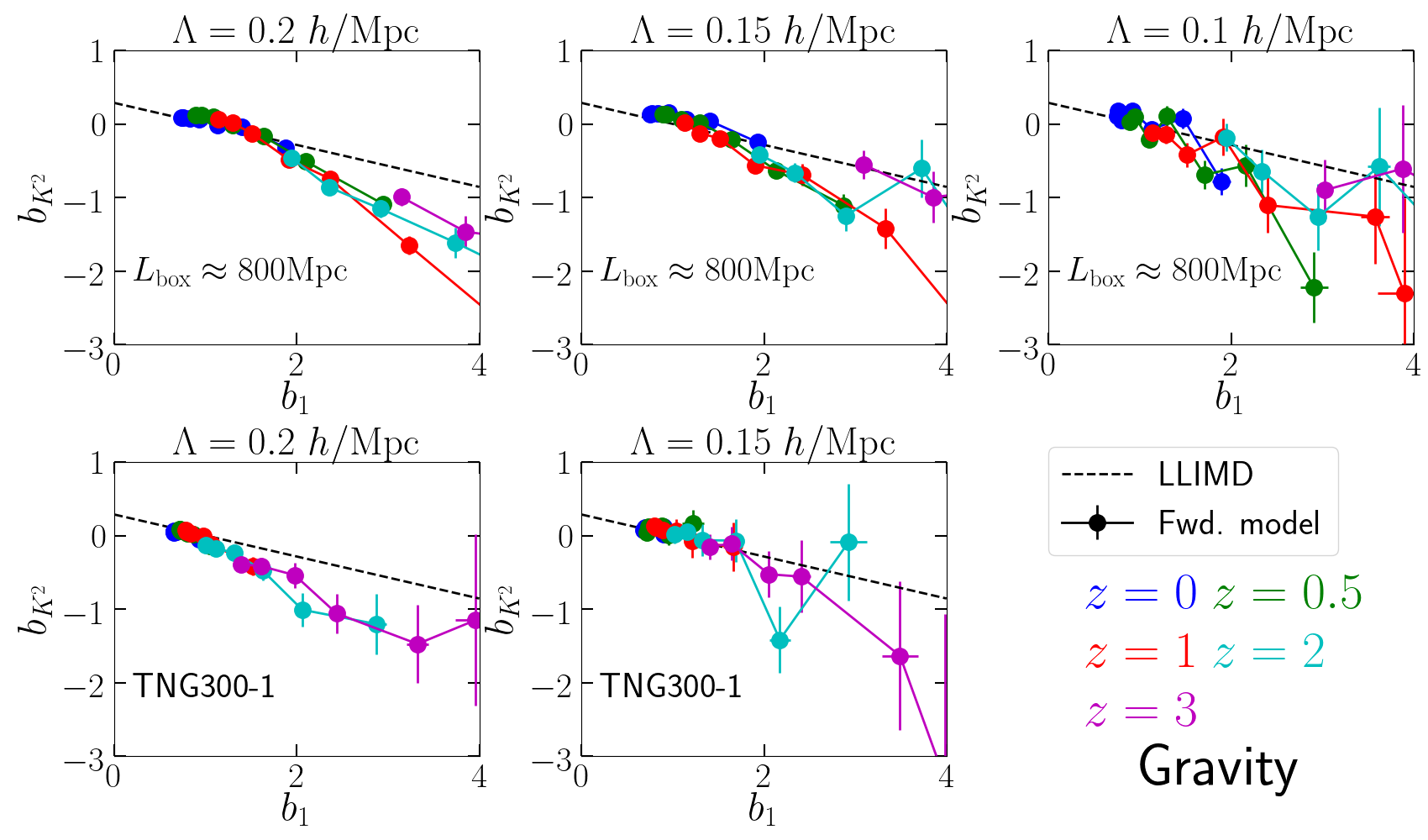}
\caption{The $b_{K^2}(b_1)$ relation of dark matter halos in the Gravity runs; this is the same as Fig.~\ref{fig:fig_biaspart_Lambda_convergence_b2_both_tng300_1_tng800_1_group_dmo_totmass}, but for $b_{K^2}$ instead of $b_2$. The black dashed line shows the LLIMD prediction \cite{sheth/chan/scoccimarro:2012, chan/scoccimarro/sheth:2012, baldauf/etal:2012, MSZ, 2018JCAP...09..008L}.}
\label{fig:fig_biaspart_Lambda_convergence_bK2_both_tng300_1_tng800_1_group_dmo_totmass}
\end{figure}

\begin{figure}
\centering
\includegraphics[width=\textwidth]{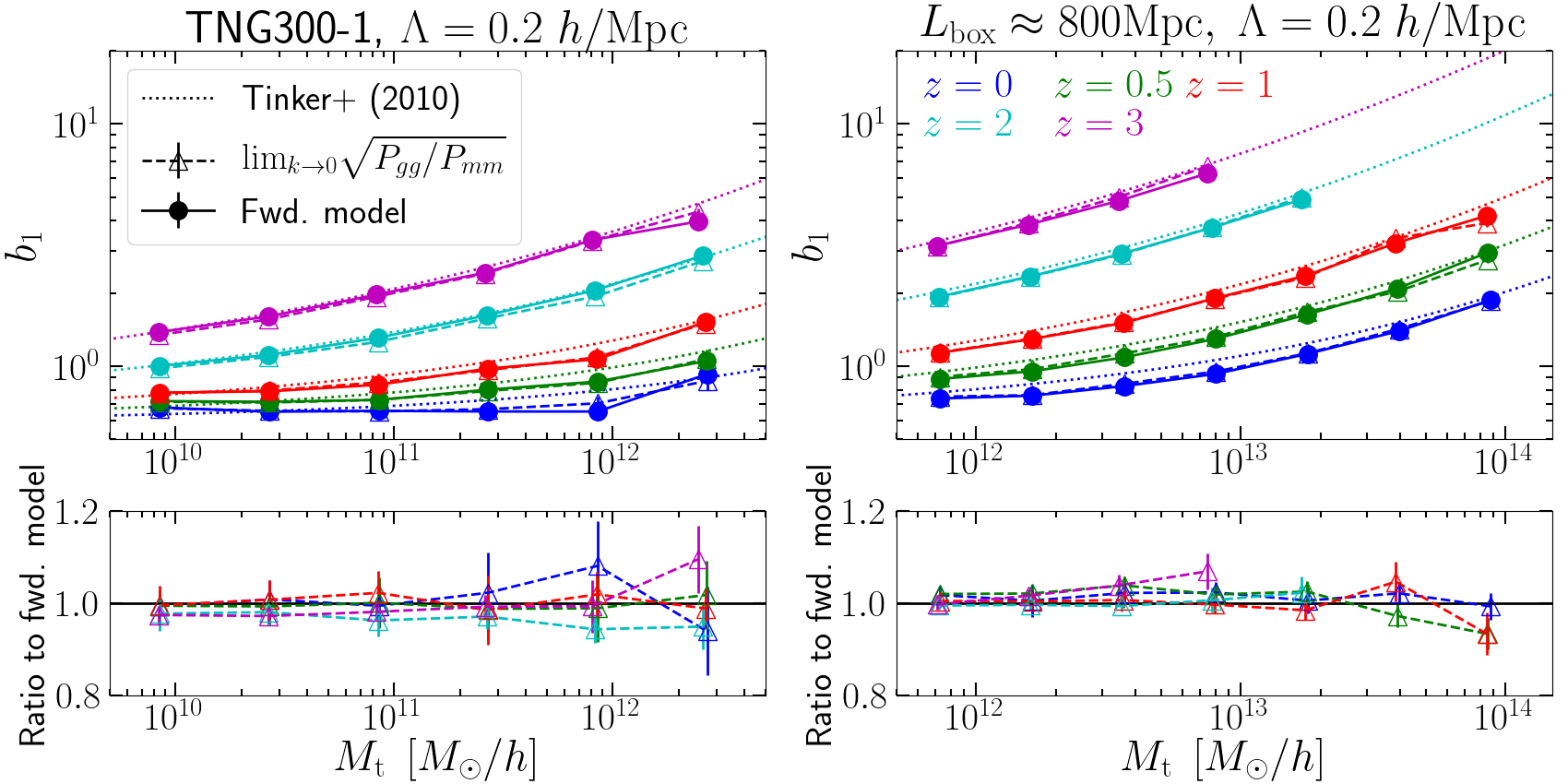}
\caption{The $b_1$ parameter of dark matter halos as a function of their total mass $M_t$ in the Gravity runs. The left and right panels are for the two boxes and the colored points show the result obtained with the forward model and the EFT likelihood of Eq.~(\ref{eq:eftlike}) with $\Lambda = 0.2h/{\rm Mpc}$ for different redshifts, as labeled. The dotted lines show the result from the fitting formula of Ref.~\cite{2010ApJ...724..878T}. The open triangles show the result obtained using the large-scale limit of the ratio of the halo to matter power spectra; the lower panels show the ratio of these to the forward model result.}
\label{fig:fig_biaspart_Lambda_convergence_0p2_b1_both_tng300_1_tng800_1_group_dmo_totmass}
\end{figure}

Before analysing the simulated galaxies from the Hydro simulation, we begin by investigating the impact of the choice of the cutoff $\Lambda$ on the $b_2(b_1)$ and $b_{K^2}(b_1)$ relations inferred for dark matter halos in the Gravity runs. This way we can compare to results from previous works in the literature to validate and build confidence about our numerical methodology.

The colored points in Fig.~\ref{fig:fig_biaspart_Lambda_convergence_b2_both_tng300_1_tng800_1_group_dmo_totmass} show the inferred $b_2(b_1)$ relation for the halos in the Gravity runs for three values of the cutoff, $\Lambda = 0.2, 0.15, 0.1 h/{\rm Mpc}$. The result is for halos selected by their total mass, i.e., each point marks the $b_2$ and $b_1$ values of halos found in the same mass bin (these mass values can be read from Fig.~\ref{fig:fig_biaspart_Lambda_convergence_0p2_b1_both_tng300_1_tng800_1_group_dmo_totmass}). The black dashed line shows the fitting formula of Ref.~\cite{lazeyras/etal} obtained with separate universe simulations, $b_2(b_1) = 0.412 - 2.143b_1 + 0.929b_1^2 + 0.008b_1^3$, which we take here as the reference result. For both the TNG300-1 and $\tngbig$ boxes, as well as all $\Lambda$ and redshift values, the result using the EFT likelihood reproduces well the expected result, which constitutes an important validation test of the forward modeling and EFT likelihood approach. A closer inspection reveals however a small trend for the forward model result to overpredict the expected $b_2$ values for $b_1 \lesssim 2$, for the two highest values of the cutoff $\Lambda = 0.2, 0.15h/{\rm Mpc}$. This overprediction becomes less noticeable when $\Lambda = 0.1h/{\rm Mpc}$, as seen in the upper right panel for the bigger box $\tngbig$. This small effect can be explained most likely as a mismatch between the bias parameters inferred by the forward model, which refer explicitly to the scale $\Lambda$, and the separate-universe bias parameters (or equivalently the bias parameters of large-scale $N$-point functions), which refer to the $k\to 0$ limit. While the dominant correction has been incorporated through the renormalization with respect to $\delta_m$, there are additional corrections involving $\delta_m^2, K^2$ themselves which are not accounted for in this paper (cf.~Sec.~\ref{sec:biasexp}). In addition, higher-order bias contributions not included in the forward model, starting from fourth order, can also systematically shift the inferred bias parameters (recall, we stop at third order in the bias expansion; cf.~Eqs.~(\ref{eq:biasexp_E}) and (\ref{eq:operators})). Since both of these corrections shrink for smaller $\Lambda$,  this would suggest adopting $\Lambda = 0.1h/{\rm Mpc}$ as the default value to obtain our results, but unfortunately, this yields too low signal-to-noise estimates of both $b_2$ and $b_{K^2}$ from the smaller TNG300-1 box (not shown): its smaller volume does not provide enough statistical power (i.e.~enough number of modes) after the sharp-$k$ filtering with $\Lambda = 0.1h/{\rm Mpc}$ is applied. For this reason, in our main results in the next section, we will default to using $\Lambda = 0.2h/{\rm Mpc}$ as it provides the highest signal-to-noise results, but we shall keep in mind this small systematic shift ($\Delta\left(b_2/2\right) \sim 0.2$) when interpreting our findings. We refer the reader to our companion paper \cite{Lazeyras:2021dar} for a more detailed inspection of the $\Lambda$-dependence of the bias parameters, including a strategy to infer the bias parameters in the limit $\Lambda \to 0$.

Figure \ref{fig:fig_biaspart_Lambda_convergence_bK2_both_tng300_1_tng800_1_group_dmo_totmass} shows the same as Fig.~\ref{fig:fig_biaspart_Lambda_convergence_b2_both_tng300_1_tng800_1_group_dmo_totmass}, but for $b_{K^2}(b_1)$ instead of $b_2(b_1)$. Here, the dashed line shows the so-called Lagrangian local-in-matter-density (LLIMD) prediction, $b_{K^2} = -(2/7)(b_1-1)$ \cite{sheth/chan/scoccimarro:2012, chan/scoccimarro/sheth:2012, baldauf/etal:2012, MSZ, 2018JCAP...09..008L}, which is obtained by neglecting the effects of the large-scale tidal field at the initial time and considering only the tidal contribution sourced by the subsequent gravitational evolution. This is only expected to yield a rough approximation to $b_{K^2}$, and so it not surprising that the EFT likelihood result does not recover the LLIMD relation. Instead, our results are in line with past estimates of $b_{K^2}$ \cite{saito/etal:14, 2018JCAP...09..008L, 2018JCAP...07..029A, 2020PhRvD.102j3530E, 2021arXiv210206902E} that generically lie below the LLIMD relation for $b_1 \gtrsim 2$. Unlike our $b_2$ results in Fig.~\ref{fig:fig_biaspart_Lambda_convergence_b2_both_tng300_1_tng800_1_group_dmo_totmass}, we do not discern in Fig.~\ref{fig:fig_biaspart_Lambda_convergence_bK2_both_tng300_1_tng800_1_group_dmo_totmass} any obvious systematic trend with the value of $\Lambda$. Perhaps the main noteworthy difference is a noticeable decrease in precision for the $\tngbig$ box with $\Lambda = 0.1h/{\rm Mpc}$, suggesting that even this bigger box begins to loose some constraining power on $b_{K^2}$ when this cutoff is employed.

For completeness, Fig.~\ref{fig:fig_biaspart_Lambda_convergence_0p2_b1_both_tng300_1_tng800_1_group_dmo_totmass} shows the dependence of $b_1$ on the total mass of the halos for $\Lambda = 0.2 h/{\rm Mpc}$. The result obtained using the forward model (points) is in very good agreement with that obtained from the large-scale limit of the ratio of the halo to matter power spectra (open triangles). We determine the latter by fitting the function $f(k) = b_1 + ak^2$ (where $b_1$ and $a$ are free coefficients) to $\sqrt{P_{gg}(k)/P_{mm}(k)}$ for $k < 0.15 h/{\rm Mpc}$. The lower panels display the ratio of these two estimates of $b_1$, where we note an agreement that is generically better than $5\%$ and within the quoted errors. The dotted lines in the upper panels of Fig.~\ref{fig:fig_biaspart_Lambda_convergence_0p2_b1_both_tng300_1_tng800_1_group_dmo_totmass} show the result of the fitting formula of Ref.~\cite{2010ApJ...724..878T}. The differences to the points and triangles are $\approx 15\%$ in the worst cases, which is in line with the expected accuracy of the fitting formula, as well as some sample variance at the higher-mass end in each simulation box. 

Overall, the successful recovery of the expected $b_2(b_1)$ and $b_{K^2}(b_1)$ relations by the forward model and the EFT likelihood is highly nontrivial and it demonstrates the ability of this approach to utilize the higher-order nonlinear information encoded in the galaxy distribution. 

\subsection{Galaxy bias as a function of simulated galaxy properties}
\label{sec:mainres}

\begin{figure}
\begin{subfigure}
\centering
\includegraphics[width=\textwidth]{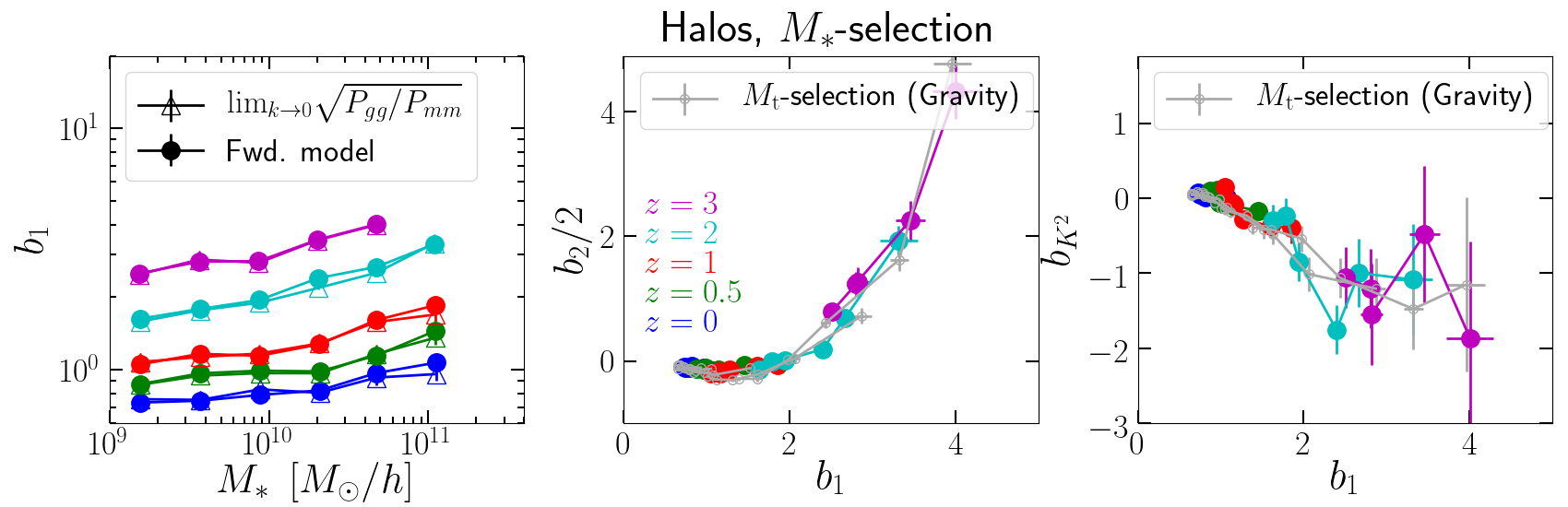}
\end{subfigure}
\begin{subfigure}
\centering
\includegraphics[width=\textwidth]{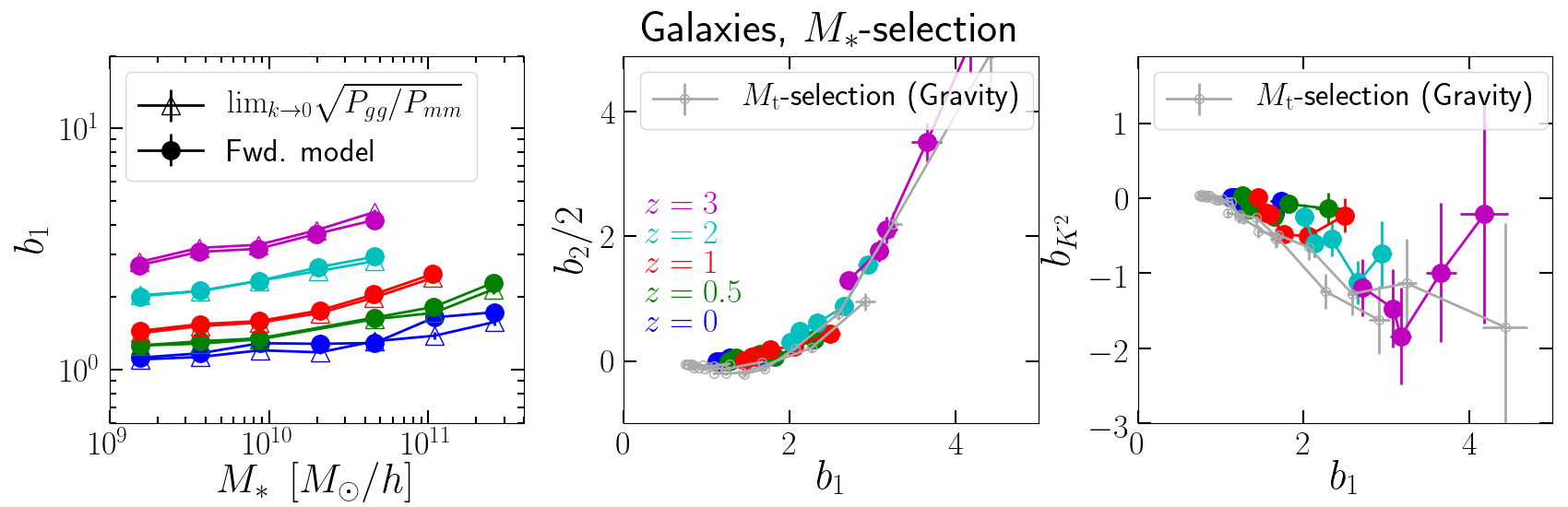}
\end{subfigure}
\caption{The bias parameters $b_1$, $b_2$ and $b_{K^2}$ for halos (top) and galaxies (bottom) selected by their stellar mass $M_*$ in the Hydro TNG300-1 simulation. The left panels show $b_1$ in stellar mass bins, and the middle and right panels show the $b_2(b_1)$ and $b_{K^2}(b_1)$ relations for the same bins, respectively. The colored points show the result obtained with the forward model and EFT likelihood of Eq.~(\ref{eq:eftlike}) for $\Lambda = 0.2h/{\rm Mpc}$ at different redshifts, as labeled; the open triangles in the left panels show the $b_1$ values obtained using the large-scale limit of the ratio of the halo/galaxy to matter power spectra. For comparison, in the middle and right panels, the grey points show the relations obtained from the Gravity simulation for total mass $M_t$ selected objects (halos on top and subhalos at the bottom); these are not distinguished by redshift to lighten the figure and we have also checked that the total-mass selected results of the Gravity and Hydro runs are nearly indistinguishable.}
\label{fig:fig_biaspart_Lambda_02_tng300_1_part1}
\end{figure}

We turn our attention now to the galaxy bias parameters estimated for objects in the Hydro run, i.e., the full-physics TNG300-1 simulation. Figure \ref{fig:fig_biaspart_Lambda_02_tng300_1_part1} shows the bias parameters $b_1$, $b_2$ and $b_{K^2}$ for stellar mass $M_*$ selected halos and galaxies. The left panels show the stellar mass dependence of $b_1$, where we see once again a very good agreement between the forward model result for $\Lambda = 0.2h/{\rm Mpc}$ (points) and the estimate based on the large-scale limit of the galaxy to matter power spectra. It is perhaps worth noticing that, as expected, at fixed $M_*$ the bias of the galaxies (bottom) is larger than the bias of the halos (top); this is because galaxies live inside halos with higher stellar mass, which are therefore more biased. We refer the reader to Refs.~\cite{2018MNRAS.475..676S, 2020MNRAS.496.1182M} for more in-depth studies of the $b_1$ values of the galaxies in the TNG300-1 simulation.

Concerning the $b_2$ results, the middle panels of Fig.~\ref{fig:fig_biaspart_Lambda_02_tng300_1_part1} show that the $b_2(b_1)$ relation obtained from the Gravity simulations for total mass $M_t$ selected objects (shown by the grey points) is fairly well reproduced even when the objects are selected by their stellar mass $M_*$ (colored points). This agreement is better for the case of halos. For the case of galaxies, a closer inspection suggests a small trend for the $b_2$ values of the stellar mass selected galaxies to be slightly larger (by about $\Delta(b_2/2)\approx 0.1-0.2$) for $b_1 \lesssim 2$, compared to the $b_2$ values of the subhalos selected by total mass. This hint is not very significant, and with just one realization of the simulations as well as the similar shift associated with $\Lambda = 0.2h/{\rm Mpc}$ discussed in the last section in Fig.~\ref{fig:fig_biaspart_Lambda_convergence_b2_both_tng300_1_tng800_1_group_dmo_totmass}, we will refrain from drawing too strong conclusions based on this result. We will discuss in the next section, however, possible physical explanations for differences between the $b_2(b_1)$ relation when the objects are selected according to different properties.

Within the larger error bars in the right panels of Fig.~\ref{fig:fig_biaspart_Lambda_02_tng300_1_part1}, we note again a broad agreement between the $b_{K^2}(b_1)$ relation of the objects selected by $M_*$ (colored) and $M_t$ (grey). Just like for $b_2$, the agreement appears better for the case of the halos, especially for $b_1 \lesssim 2$ where the error bars are smaller. For stellar mass selected galaxies, on this $b_1$ range, there is again a small hint for larger $b_{K^2}$ values ($\Delta(b_{K^2})\approx 0.2$) compared to total mass selection, which is worth confirming in future work with more realizations or larger volumes of galaxy formation simulations. Our $b_{K^2}$ results in Fig.~\ref{fig:fig_biaspart_Lambda_02_tng300_1_part1} show also an intriguing feature on the $b_{K^2}(b_1)$ relation of the stellar mass selected objects at $z=2$ and $z=3$, which exhibits a non-monotonic dependence on $b_1$ that varies steeply in some $b_1$ intervals; this is especially noticeable for the $z=3$ galaxy results (lower right panel). With just a single, relatively small-volume simulation, it is hard to rule out this as due to numerical noise, but we mention it explicitly anyway in light of the results in our companion paper \cite{Lazeyras:2021dar} where similar (and higher signal-to-noise) features are also observed in halo assembly bias, i.e. when halos in gravity-only (larger-volume) simulations are selected by properties other than their total mass (namely concentration, spin and sphericity). Given the precision of our measurements here, we opt to be careful in drawing decisive conclusions, but note this is worth clarifying in future work.

\begin{figure}
\begin{subfigure}
\centering
\includegraphics[width=\textwidth]{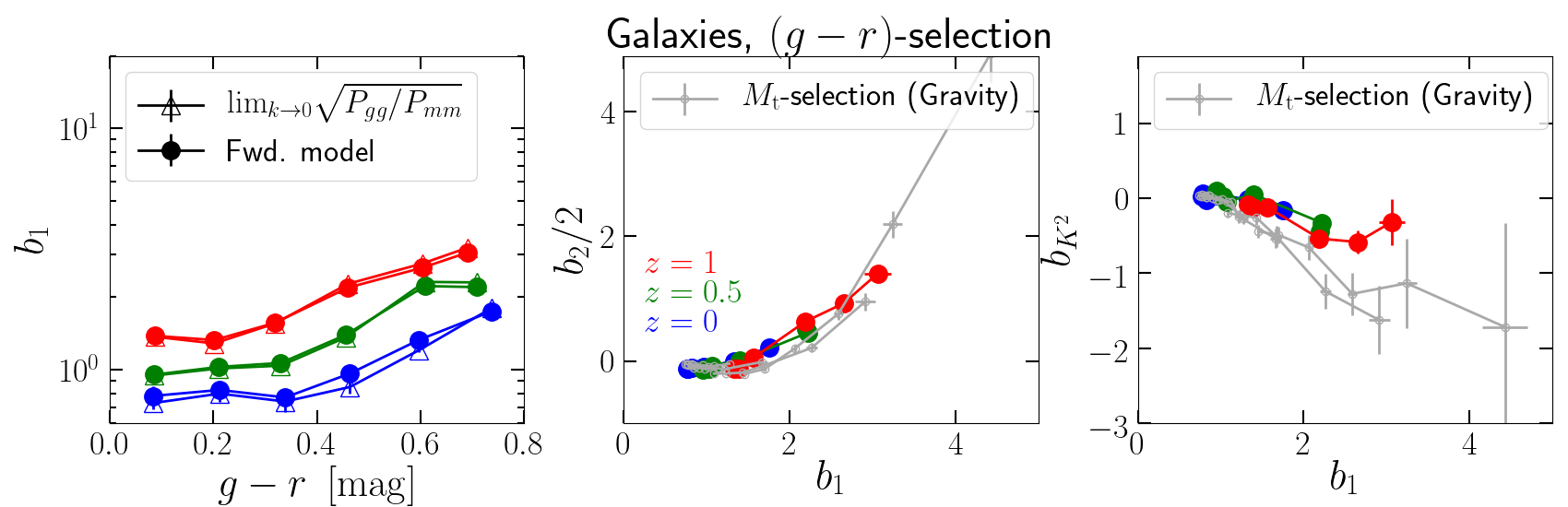}
\end{subfigure}
\begin{subfigure}
\centering
\includegraphics[width=\textwidth]{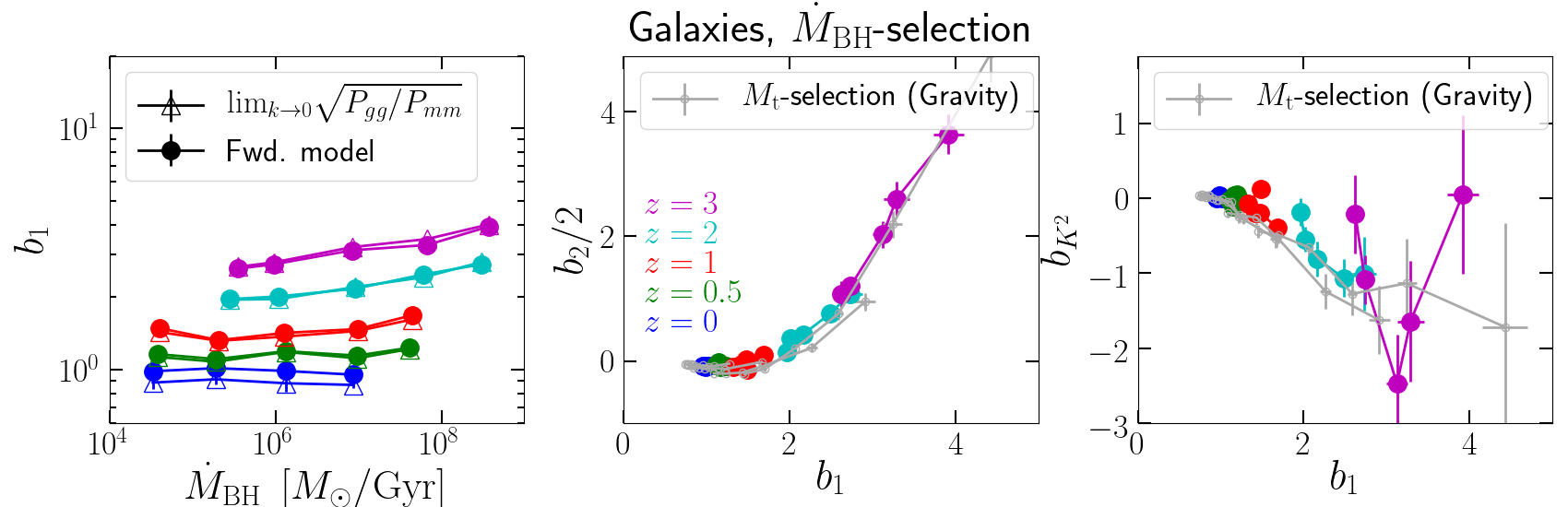}
\end{subfigure}
\begin{subfigure}
\centering
\includegraphics[width=\textwidth]{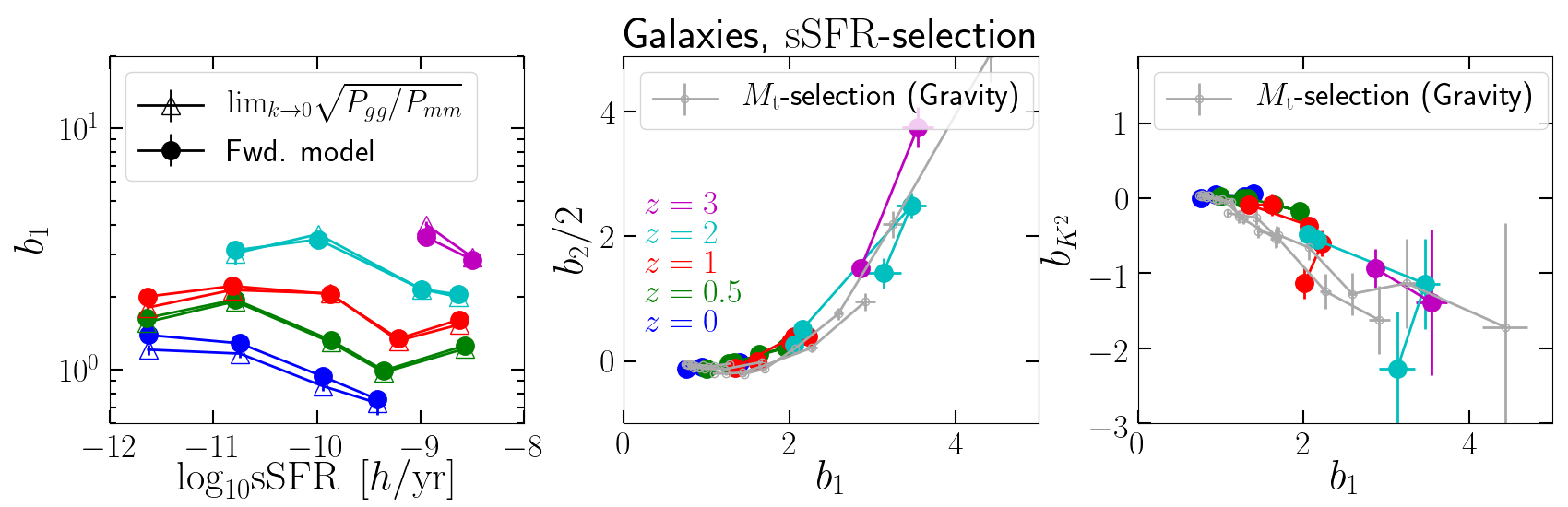}
\end{subfigure}
\caption{The bias parameters $b_1$, $b_2$ and $b_{K^2}$ for galaxies selected by $(g-r)$ color (top), black hole accretion rate $\dot{M}_{\rm BH}$ (center) and specific star formation rate ${\rm sSFR}$ (bottom) in the Hydro TNG300-1 simulation. These panels show the same as the bottom ones in Fig.~\ref{fig:fig_biaspart_Lambda_02_tng300_1_part1}, but for these galaxy selection criteria, instead of stellar mass $M_*$. Note that galaxies in bins of each of these three properties span a wide range in stellar mass.}
\label{fig:fig_biaspart_Lambda_02_tng300_1_part2}
\end{figure}

Figure \ref{fig:fig_biaspart_Lambda_02_tng300_1_part2} shows our $b_1$, $b_2$ and $b_{K^2}$ results for galaxies selected by $(g-r)$ color (note, without dust correction \cite{Nelson:2017cxy}), black hole accretion rate $\dot{M}_{\rm BH}$ and specific star formation rate ${\rm sSFR}$ (star formation rate per unit stellar mass). The $b_1$ results are in line with what one would expect physically. For example, (i) red galaxies (larger $g-r$) tend to be also the most massive ones at $z<1$ (cf.~Fig.~\ref{fig:fig_sigma8part_scatter}), and so have larger $b_1$; (ii) $b_1$ grows slightly with $\dot{M}_{\rm BH}$ at $z>2$, reflecting the fact that the black holes in massive galaxies at earlier times accrete faster (this becomes less pronounced at later times as the correlation between mass and $\dot{M}_{\rm BH}$ becomes weaker); and (iii) the fraction of quenched galaxies increases with halo mass, which explains why $b_1$ decreases with ${\rm sSFR}$, i.e., strongly star forming galaxies tend to be smaller, less biased galaxies (cf.~Fig.~\ref{fig:fig_sigma8part_scatter}). Note that galaxies in bins of each of these three properties cover a large range of total and stellar mass values, and thus correspond to very different galaxy populations with different total number densities. We have explicitly checked (not shown) that the $b_1$ values also depend markedly on the adopted selection criteria, even at fixed galaxy number density (similarly to Fig.~9 of Ref.~\cite{2014MNRAS.442.2131A} and Fig.~18 of Ref.~\cite{2018MNRAS.475..676S}).

Regarding the $b_2$ results in Fig.~\ref{fig:fig_biaspart_Lambda_02_tng300_1_part2}, we note again a very good agreement between the $b_2(b_1)$ relations of the total mass selected subhalos in the Gravity run (grey) and that of the galaxies. Contrary to the case of stellar mass selection, here we do not discern any hint for larger $b_2$ for $b_1 \lesssim 2$. Interestingly, for $b_{K^2}$, our results show that the values for $(g-r)$ selected galaxies are larger ($\Delta(b_{K^2})\approx 0.5$) than those for total mass selection for $b_1 \gtrsim 1.5$; the next section discusses possible physical interpretations of differences like this. Further, the {\it sharp, non-monotonic} features of the $b_{K^2}(b_1)$ relation that we mentioned above are especially noticeable in the case of galaxies selected by $\dot{M}_{\rm BH}$ (right center panel); this strengthens the motivation for follow up studies with more realizations and (if possible) larger volume galaxy formation simulations to clarify the origin of this result and any eventual ties to the galaxy assembly bias signal (cf.~our companion paper Ref.~\cite{Lazeyras:2021dar}).

\subsection{Halo model interpretation of galaxy bias relations}
\label{sec:hm}

The halo model \cite{cooray/sheth} and halo occupation distribution (HOD) formalisms \cite{1997MNRAS.286..795K, 2000MNRAS.318..203S, 2000MNRAS.318.1144P, 2003ApJ...593....1B, 2004ApJ...609...35K, 2020arXiv201204637V} can be used to interpret some of our galaxy bias results in the last section. The starting assumption is that galaxies live inside dark matter halos of some total mass $M_{\rm h} \equiv M_t^{\rm h}$, and so the number density of galaxies with total mass $M_t^{g}$ and stellar mass $M_*^{g}$, can be written, respectively, as
\bq\label{eq:ng}
n_g(M_t^{g}) &=& \int {\rm d}M_{\rm h} n_{\rm h}(M_{\rm h}) N_g(M_t^{g}|M_{\rm h}) \nonumber \\
n_g(M_*^{g}) &=& \int {\rm d}M_{\rm h} n_{\rm h}(M_{\rm h}) N_g(M_*^{g}|M_{\rm h}),
\eq
where $n_{\rm h}$ is the halo mass function and $N_g(M^{g}|M_{\rm h})$ is the HOD number, i.e., the mean number of galaxies with mass $M^{g}$ that live in halos with mass $M_{\rm h}$. The two HOD numbers above can be related as
\bq\label{eq:HOD}
N_g(M_*^{g}|M_{\rm h}) =  \int{\rm d}M_t^{g}  N_g(M_t^{g}|M_{\rm h}) \mathcal{P}(M_*^g|M_t^g),
\eq
where $\mathcal{P}(M_*^g|M_t^g)$ is the galaxy stellar-to-total-mass relation, i.e., the probability that a galaxy with stellar mass $M_*^g$ has total mass $M_t^g$. In this discussion, we focus specifically on total and stellar masses, but this generalizes straightforwardly to any other choice of galaxy properties.

To build intuition for the second-order bias parameters, let us inspect first the case of the linear LIMD parameter $b_1$, which is defined as the first-order response of the galaxy number density to long-wavelength total mass perturbations, i.e., $b_1 = (\partial n_g / \partial \delta_m)/n_g$. Using the expressions above, for total galaxy mass $M_t^g$ selection we have
\bq\label{eq:b1_totmass}
b_1(M_t^{g}) &=& \frac{1}{n_g(M_t^{g})} \int {\rm d}M_{\rm h} \ n_{\rm h}(M_{\rm h}) N_g(M_t^{g}|M_{\rm h}) \left(b_1^{\rm h}(M_{\rm h})  +  R_1^{N_g}(M_t^{g}|M_{\rm h})\right),
\eq
where $b_1^{\rm h} = (\partial n_{\rm h} / \partial \delta_m)/n_h$ is the linear bias of the halos and $R_1^{N_g}$ is the first-order response of the HOD number of total-mass selected galaxies, i.e., $R_1^{N_g} = (\partial N_g / \partial \delta_m)/N_g$.  If $R_1^{N_g} = 0$, then one recovers the popular interpretation of galaxy bias as a HOD-weighted version of the bias of the dark matter halos; see however Ref.~\cite{2020arXiv201204637V} for a recent study of the response functions of HOD numbers and their corresponding impact on galaxy bias predictions. For the case of stellar mass selection, we have instead
\bq\label{eq:b1_stemass}
b_1(M_*^{g}) &=& \frac{1}{n_g(M_*^{g})} \int {\rm d}M_{\rm h} \int {\rm d}M_t^g\ n_{\rm h}(M_{\rm h}) N_g(M_t^{g}|M_{\rm h}) \mathcal{P}(M_*^g|M_t^g)  \nonumber \\
&& \times \left(b_1^{\rm h}(M_{\rm h})  +  R_1^{Ng}(M_t^{g}|M_{\rm h}) + R_1^{\mathcal{P}}(M_*^g|M_t^g)\right),
\eq
which depends additionally on the first-order response of the galaxy stellar-to-total-mass relation $R_1^{\mathcal{P}} = (\partial\mathcal{P}/ \partial \delta_m)/\mathcal{P}$. Comparing Eqs.~(\ref{eq:b1_totmass}) and (\ref{eq:b1_stemass}) tells us that if the galaxy stellar-to-total-mass relation does not depend to first-order on the large-scale overdensity, i.e., $R_1^{\mathcal{P}}(M_*^g|M_t^g) = 0$, then the bias of the stellar mass selected galaxies can be obtained from that of total mass selected galaxies via a simple mapping of $M_t^g$ to $M_*^g$ given by $\mathcal{P}(M_*^g|M_t^g)$. If on the other hand, the galaxy stellar-to-total-mass relation is sensitive to whether the galaxies live in large-scale overdense or underdense regions, then there is an additional contribution coming from $R_1^{\mathcal{P}}(M_*^g|M_t^g) \neq 0$ (see Refs.~\cite{2020JCAP...12..013B, 2020arXiv201204637V} for a discussion and practical demonstrations of the importance of these response functions).

The second-order LIMD bias parameter is defined as $b_2 = (\partial^2 n_g / \partial \delta_m^2)/n_g$, and the corresponding expressions follow straightforwardly. For total mass selection, we have
\bq\label{eq:b2_totmass}
b_2(M_t^{g}) &=& \frac{1}{n_g(M_t^{g})} \int {\rm d}M_{\rm h} \ n_{\rm h}(M_{\rm h}) N_g(M_t^{g}|M_{\rm h}) \left(b_2^{\rm h}(M_{\rm h})  + 2b_1^{\rm h}(M_{\rm h})R_1^{N_g}(M_t^{g}|M_{\rm h}) +  R_2^{N_g}(M_t^{g}|M_{\rm h})\right), \nonumber \\
\eq
where $b_2^{\rm h}(M_{\rm h})$ is the second-order LIMD bias of the halos, and for stellar mass selection we have
\bq\label{eq:b2_stemass}
b_2(M_*^{g}) &=& \frac{1}{n_g(M_*^{g})} \int {\rm d}M_{\rm h} \int {\rm d}M_t^g\ n_{\rm h}(M_{\rm h}) N_g(M_t^{g}|M_{\rm h}) \mathcal{P}(M_*^g|M_t^g)  \nonumber \\
&& \times \bigg(b_2^{\rm h}(M_{\rm h})  + 2b_1^{\rm h}(M_{\rm h})R_1^{N_g}(M_t^{g}|M_{\rm h}) +  R_2^{N_g}(M_t^{g}|M_{\rm h})   \nonumber \\
&&  \ \ \ + 2\left[b_1^{\rm h}(M_{\rm h}) + R_1^{N_g}(M_t^{g}|M_{\rm h})\right]R_1^{\mathcal{P}}(M_*^g|M_t^{g}) + R_2^{\mathcal{P}}(M_*^g|M_t^{g})\bigg),
\eq
where $R_2^{\mathcal{P}} = (\partial^2\mathcal{P}/ \partial \delta_m^2)/\mathcal{P}$ is the second-order response of the galaxy stellar-to-total-mass relation.

\begin{figure}
\includegraphics[width=\textwidth]{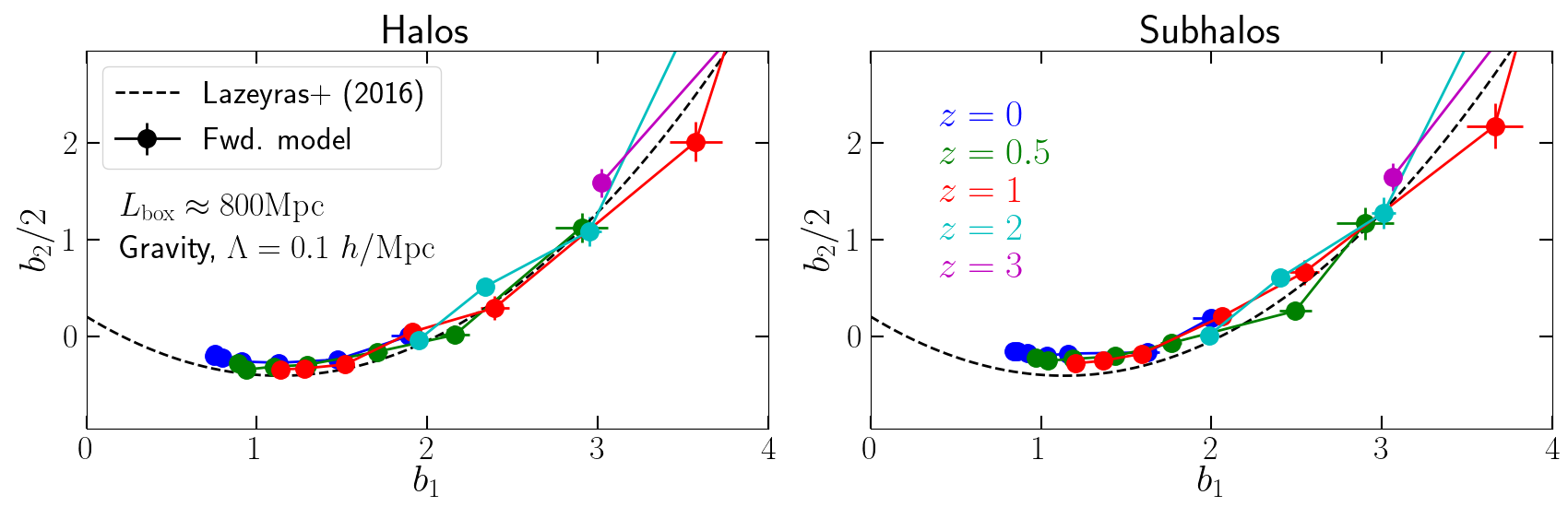}
\caption{The $b_2(b_1)$ relation for total mass selected dark matter halos (left) and subhalos (right). The colored points show the result obtained with the forward model and the EFT likelihood of Eq.~(\ref{eq:eftlike}) at different redshifts, as labeled. The result is shown only for the gravity-only $\tngbig$ box and for $\Lambda = 0.1h/{\rm Mpc}$. The black dashed line shows the fitting formula obtained by Ref.~\cite{lazeyras/etal} for halos using separate universe simulations.}
\label{fig:fig_biaspart_Lambda_convergence_0p1_b2_tng800_2_subhalo_group_dmo_totmass}
\end{figure}

Equations (\ref{eq:b1_totmass})-(\ref{eq:b2_stemass}) encode two results that we find interesting to highlight. The first concerns the $b_2(b_1)$ relations of halos and subhalos (or total mass selected galaxies), which Eqs.~(\ref{eq:b1_totmass}) and (\ref{eq:b2_totmass}) show is in general different if $R_1^{N_g} \neq 0$ or $R_2^{N_g} \neq 0$. However, even if $R_1^{N_g} = R_2^{N_g}= 0$, the nonlinearity of the $b_2(b_1)$ relation of the halos forcibly implies a different relation for the subhalos as well. Concretely, the fit from Ref.~\cite{lazeyras/etal} for halos that we show in Fig.~\ref{fig:fig_biaspart_Lambda_convergence_b2_both_tng300_1_tng800_1_group_dmo_totmass} is of the form $b_2^{\rm h} = A + B b_1^{\rm h} + C (b_1^{\rm h})^2 + D (b_1^{\rm h})^3$, which if we plug into Eq.~(\ref{eq:b2_totmass}) yields (ignoring the contribution from $R_1^{N_g}$ and $R_2^{N_g}$)
\bq\label{eq:b2_stemass_2}
b_2(M_t^{g}) &=& A + B b_1(M_t^{g}) + \frac{1}{n_g(M_t^{g})} \int {\rm d}M_{\rm h} \ n_{\rm h}(M_{\rm h}) N_g(M_t^{g}|M_{\rm h}) \left(C[b_1^{\rm h}(M_{\rm h})]^2 + D[b_1^{\rm h}(M_{\rm h})]^3\right), \nonumber \\
\eq
i.e., the relation of total mass selected halos and subhalos is only the same if $C=D=0$. Figure~\ref{fig:fig_biaspart_Lambda_convergence_0p1_b2_tng800_2_subhalo_group_dmo_totmass} compares the $b_2(b_1)$ relation of total mass selected halos (left) and subhalos (right), where it is possible to discern that the $b_2$ values of the subhalos are indeed slightly larger compared to the halos for $b_1 \lesssim 2$. This is as one would expect from the above equation since $C > 0$ and $D>0$; more generally, this follows from the convexity of the $b_2(b_1)$ relation. Reference \cite{2021arXiv210206902E} found a similar result using analyses of the power spectrum and bispectrum of halo and HOD catalogues.

The second result has to do with the effects of the responses of the galaxy stellar-to-total-mass relation, $R_1^{\mathcal{P}}$ and $R_2^{\mathcal{P}}$. If these are equal to zero, then as discussed above, the values of both $b_1$ and $b_2$ for stellar mass selected objects are obtained from those for total mass selection via a simple mapping of stellar to total mass given by $\mathcal{P}(M_*^g|M_t^g)$, which importantly, preserves the shape of the $b_2(b_1)$ relation. On the other hand, if $R_1^{\mathcal{P}} \neq 0$ or $R_2^{\mathcal{P}} \neq 0$, i.e., the galaxy stellar-to-total-mass relation depends on local large-scale overdensities, then the $b_2(b_1)$ relation is not in general preserved as $b_1(M_*^g)$ and $b_2(M_*^g)$ are affected differently by $R_1^{\mathcal{P}}$ and $R_2^{\mathcal{P}}$ (cf.~Eqs.~(\ref{eq:b1_stemass}) and (\ref{eq:b2_stemass})). In other words, any significant difference (or lack thereof) between the colored (for stellar mass selection) and grey points (for total mass selection) in Fig.~\ref{fig:fig_biaspart_Lambda_02_tng300_1_part1} can be interpreted as an indirect hint (or lack thereof) for nonzero $R_1^{\mathcal{P}}$ or $R_2^{\mathcal{P}}$. Indeed, we noted in the last section that there is a small trend in the bottom center panel of Fig.~\ref{fig:fig_biaspart_Lambda_02_tng300_1_part1} for larger values of $b_2$ for the stellar mass selected galaxies, compared to the total mass selection case at fixed $b_1$, which from this discussion could be interpreted as evidence for the impact of the response functions of the stellar-to-total mass relation. We acknowledge that more work is needed to corroborate this physical picture, which can involve running separate universe simulations of galaxy formation \cite{2019MNRAS.488.2079B, 2020JCAP...12..013B} to measure $R_1^{\mathcal{P}}$ and $R_2^{\mathcal{P}}$ directly; we defer this to future work. 

We finish this section by stressing that although we focused this discussion on $b_2(b_1)$ and stellar mass as the galaxy property, the above derivations can be straightforwardly generalized to the $b_{K^2}(b_1)$ relation, in which case we would deal with responses to tidal fields, as well as to responses of the relation between other galaxy properties and total mass.

\subsection{Comparison to bias relations inferred from observations}
\label{sec:observations}

\begin{figure}
\includegraphics[width=\textwidth]{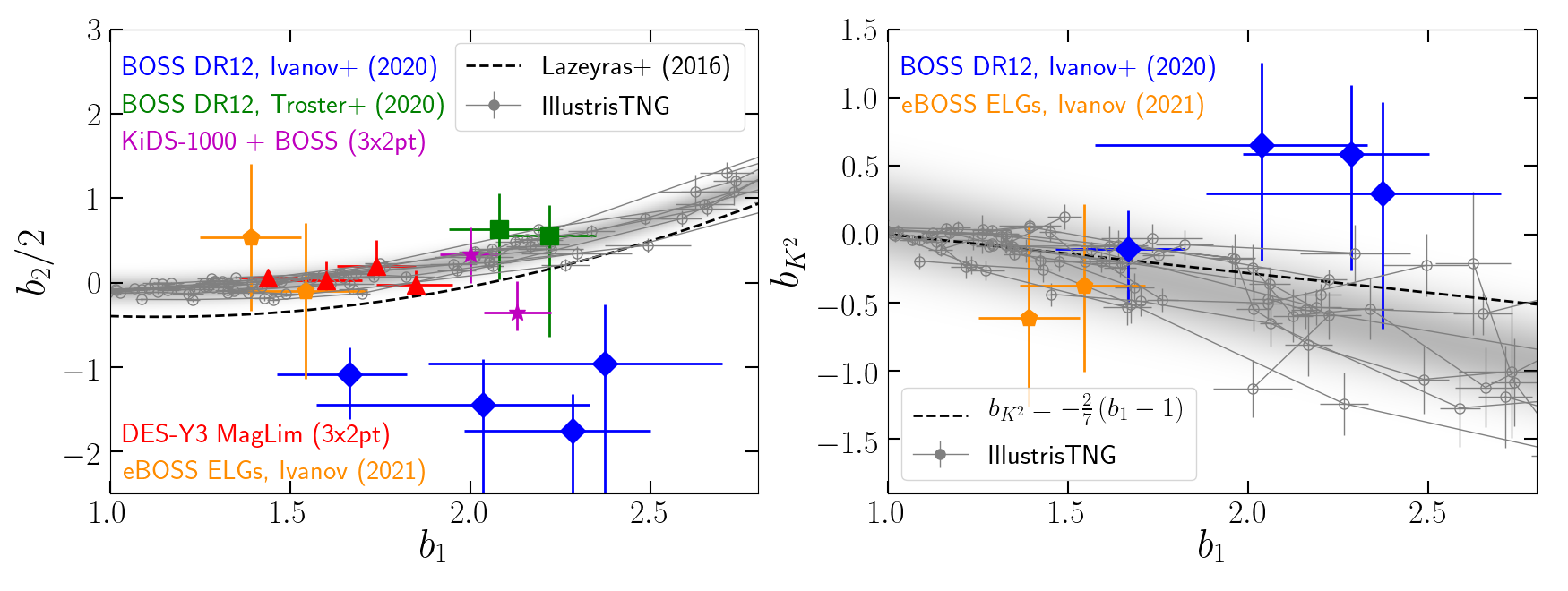}
\caption{Comparison between the $b_2(b_1)$ and $b_{K^2}(b_1)$ relations obtained in this paper for IllustrisTNG galaxies and recent estimates from real galaxy data analyses. The grey data points show the result for all of the simulated galaxy samples in Figs.~\ref{fig:fig_biaspart_Lambda_02_tng300_1_part1} and \ref{fig:fig_biaspart_Lambda_02_tng300_1_part2} put together, without distinguishing by redshift or selection criterion. The grey shaded map shows a Gaussian prior based on the IllustrisTNG data points (cf.~Eqs.~(\ref{eq:fits}) and (\ref{eq:std})). The colored data points show the relations inferred from recent cosmological inference analyses of observed galaxies: blue diamonds from Ref.~\cite{2020JCAP...05..042I}, green squares from Ref.~\cite{2020A&A...633L..10T}, magenta stars from Ref.~\cite{2021A&A...646A.140H}, red triangles from Ref.~\cite{2021arXiv210513549D}, and orange pentagons from Ref.~\cite{2021arXiv210612580I}. All error bars are $68\%$ confidence limits.}
\label{fig:obs}
\end{figure}

Figure \ref{fig:obs} compares the $b_2(b_1)$ and $b_{K^2}(b_1)$ relations obtained in this paper for IllustrisTNG galaxies (grey data points, without distinguishing by selection criteria and redshift) with recent estimates obtained from cosmological inference analyses of real galaxy samples (colored data points). The blue diamonds show the constraints obtained for the four BOSS DR12 samples (low-$z$ and high-$z$, and north and south galactic caps) analysed in Ref.~\cite{2020JCAP...05..042I} (cf.~their Table 10). Compared to our galaxy bias expansion, Ref.~\cite{2020JCAP...05..042I} replaces the $b_{K^2}K_{ij}^2$ term with $b_{\mathcal{G}_2}\mathcal{G}_2$, where $\mathcal{G}_2 = K_{ij}^2 - (2/3)\delta_m^2$. Their values of $b_2$ must therefore be subtracted by $(-4/3)b_{\mathcal{G}_2}$ before they can be compared with our simulation measurements (note also that $b_{K^2} = b_{\mathcal{G}_2}$); see App.~C of Ref.~\cite{biasreview} for the relations between different bias conventions. The green squares in Fig.~\ref{fig:obs} are the constraints obtained for the BOSS DR12 low-$z$ and high-$z$ samples in Ref.~\cite{2020A&A...633L..10T} (cf.~their Table B.1), the magenta stars show the result also for the BOSS DR12 low-$z$ and high-$z$ samples, but from the 3x2pt analyses in Ref.~\cite{2021A&A...646A.140H} using also lens galaxies from KiDS (we use the values in the column "3x2pt (joint)" in their Table C.1), and the orange pentagons are the constraints for the eBOSS ELG sample from Ref.~\cite{2021arXiv210612580I}. The second-order operators of Refs.~\cite{2020A&A...633L..10T, 2021A&A...646A.140H, 2021arXiv210612580I} are the same as in Ref.~\cite{2020JCAP...05..042I}, and so we apply the same shift on $b_2$. Finally, the red triangles show the constraints obtained for the 4 tomographic bins of the MagLim sample used in the 3x2pt analyses of DES, and that are quoted in Table V of Ref.~\cite{2021arXiv210513549D} and adopt the same second-order operators that we do here \cite{2021arXiv210513548K}.

The grey shaded map shows the amplitude of a Gaussian prior on the $b_2(b_1)$ and $b_{K^2}(b_1)$ relations based on the IllustrisTNG results. Concretely, the mean values are given by the polynomial fits 
\bq\label{eq:fits}
b_2^{\rm fit}(b_1) &=& 0.30 - 0.79b_1 + 0.20b_1^2 + 0.12b_1^3, \\ \nonumber
b_{K^2}^{\rm fit}(b_1) &=& 0.66 - 0.57b_1,
\eq
which were obtained using all IllustrisTNG data points in the range $b_1 \in \left[1, 3\right]$. The variance is assumed $b_1$-independent for simplicity, and it is given by the standard deviation of the measured bias parameters around the fits; for both $b_2$ and $b_{K^2}$ the values are 
\bq\label{eq:std}
\sigma_{b_2} = \sigma_{b_K^2} = 0.22.
\eq

Despite the very different nature of the simulated and real galaxy samples, we find the level of agreement between their bias relations in Fig.~\ref{fig:obs} to be satisfactory overall. Concerning the $b_2(b_1)$ relation, the agreement between the simulations and the observational inferences is especially good for the DES analysis \cite{2021arXiv210513549D} (red) and the BOSS analyses of Ref.~\cite{2020A&A...633L..10T} (green). The BOSS results from Ref.~\cite{2020JCAP...05..042I} (blue) appear as outliers with respect to both the simulations and the other observational analyses, but the authors note that their analysis does not provide sufficient constraining power on the $b_2$ parameter, whose bounds are prior-dominated. Note also that the analyses of Ref.~\cite{2020JCAP...05..042I} constraints the parameter combination $b_\O\mathcal{A}_s^{1/2}$, which we convert to bounds on $b_\O$ using the quoted constraints on the primordial scalar power spectrum amplitude $\mathcal{A}_s$. On the other hand, the analyses of Refs.~\cite{2020JCAP...05..042I, 2021arXiv210612580I} are the only that fitted for $b_{K^2}$ (the others kept this parameter fixed to the LLIMD relation). In this case, there is a trend for the constraints from Ref.~\cite{2020JCAP...05..042I} (blue) to lie above the $b_{K^2}(b_1)$ relation of the IllustrisTNG galaxies, but the differences are not too significant given the error bars. Note also the constraints on the $b_2$ and $b_{K^2}$ parameters do exhibit some correlation in Ref.~\cite{2020JCAP...05..042I} (see their Fig.~11), and so the $b_{K^2}$ constraints may also be affected by the lack of constraining power on $b_2$.

A thorough comparison between theory and observations is beyond the scope of the present paper and it would benefit from more work both on the theory and observational analysis sides. Here, for the time being, we limit ourselves to interpreting Fig.~\ref{fig:obs} as an encouraging illustration of the future prospects to either (i) use simulation-based priors for these bias relations to improve cosmological constraints, or conversely, (ii) use the observational estimates themselves to test galaxy formation, given theoretical predictions from other galaxy formation models, as well as matching selection strategies applied on the simulated and real galaxy data. 

\section{Results on $\sigma_8$ inference}
\label{sec:s8results}

\begin{figure}
\includegraphics[width=\textwidth]{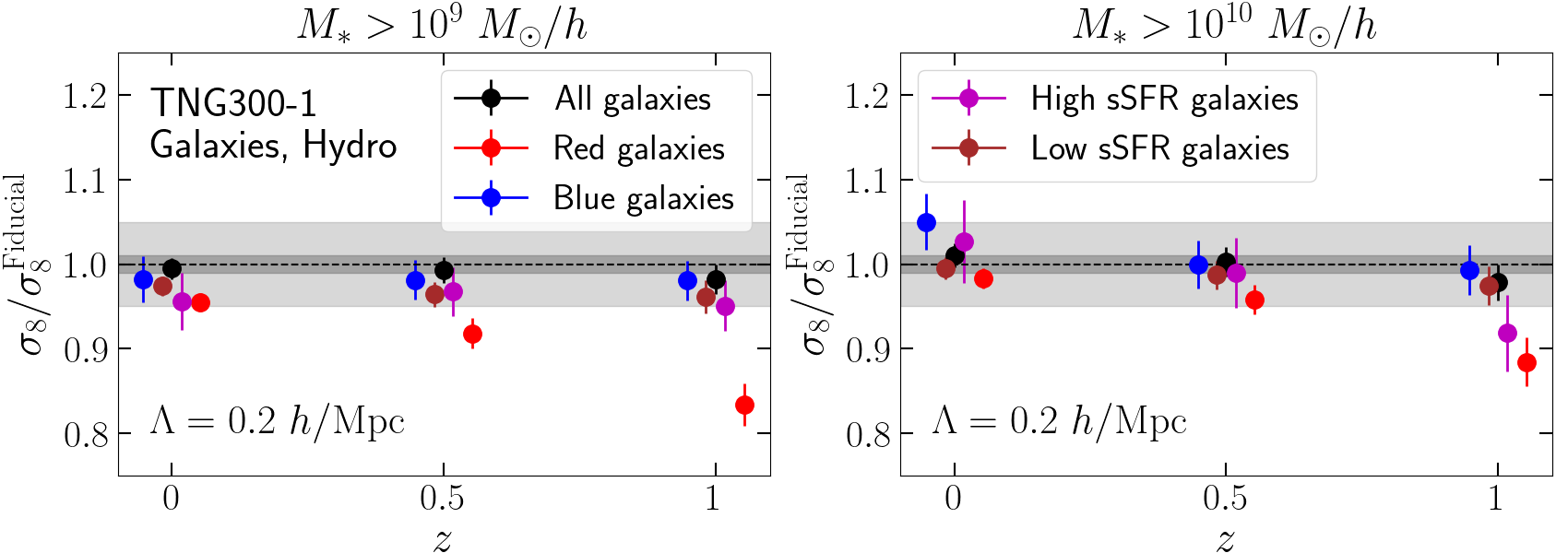}
\caption{Inferred values of $\sigma_8$ obtained using the forward model and the EFT likelihood of Eq.~(\ref{eq:eftlike}) applied (with $\Lambda = 0.2h/{\rm Mpc}$) to galaxy samples from the Hydro TNG300-1 simulation. The two panels are for different minimum stellar mass cuts, and the colors indicate different galaxy selection strategies, as labeled. The selection into red/blue and high/low ${\rm sSFR}$ galaxies is that shown in Fig.~\ref{fig:fig_sigma8part_scatter}. The two horizontal grey bands mark the $1\%$ and $5\%$ intervals. The various results at a given redshift are displaced slightly horizontally to facilitate the visualization.}
\label{fig:fig_sigma8part_Lambda_convergence_tng300_1_subhalo_hydro_stemass_splits_red}
\end{figure}

\begin{figure}
\begin{subfigure}
\centering
\includegraphics[width=\textwidth]{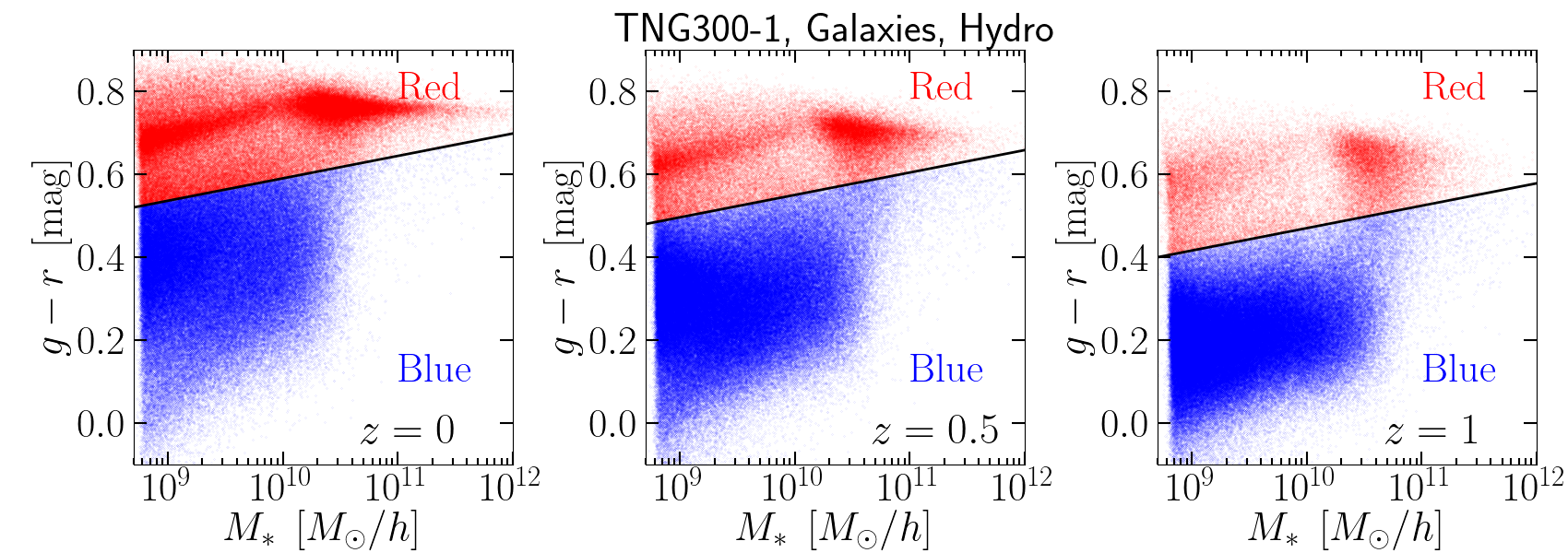}
\end{subfigure}
\begin{subfigure}
\centering
\includegraphics[width=\textwidth]{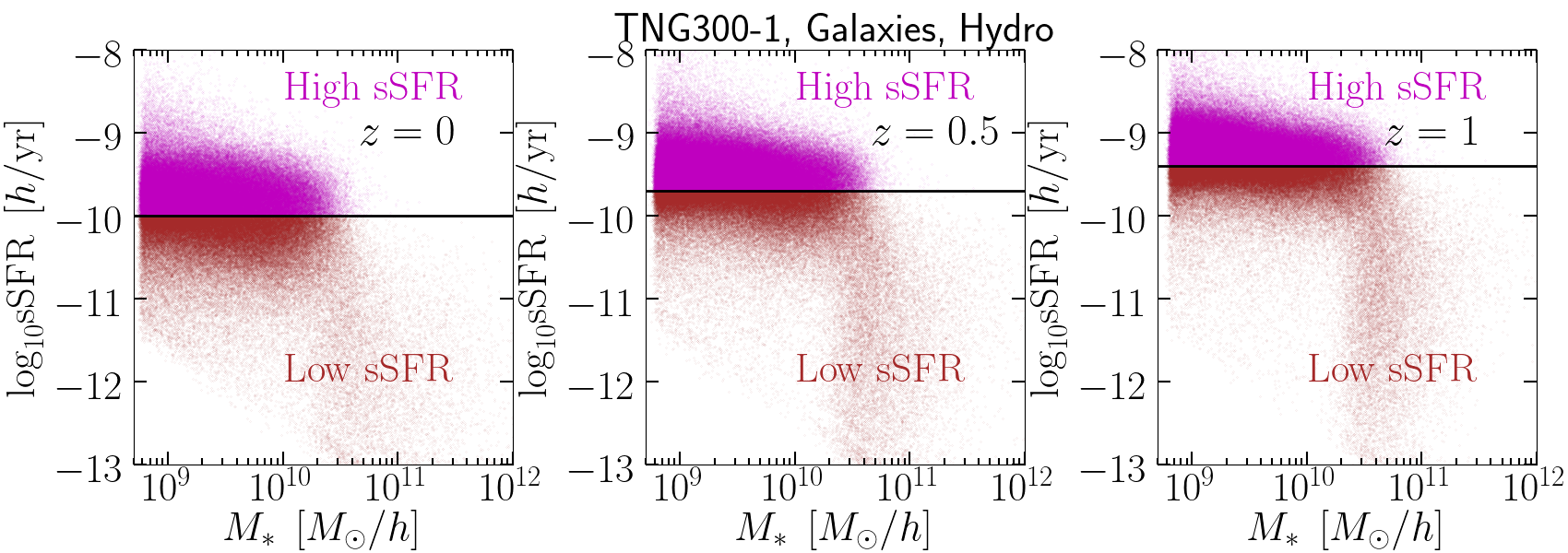}
\end{subfigure}
\caption{Distribution of the TNG300-1 galaxies in the $(g-r)-M_*$ and ${\rm sSFR}-M_*$ planes, at different redshifts as labeled. The splits into red/blue and high/low ${\rm sSFR}$ galaxies indicate the galaxy samples used in the $\sigma_8$ inference shown in  Fig.~\ref{fig:fig_sigma8part_Lambda_convergence_tng300_1_subhalo_hydro_stemass_splits_red}.}
\label{fig:fig_sigma8part_scatter}
\end{figure}

Before concluding, we show in this section the results on the $\sigma_8$ inference obtained using the forward modeling approach with the EFT likelihood of Eq.~(\ref{eq:eftlike}). The work of Refs.~\cite{2020JCAP...11..008S, 2020arXiv200914176S} has presented an in-depth study of these inference analyses using halos as tracers; here we simply apply the same formalism to various simulated galaxy samples instead. The inferred values of $\sigma_8$ are shown in Fig.~\ref{fig:fig_sigma8part_Lambda_convergence_tng300_1_subhalo_hydro_stemass_splits_red} for different redshifts and samples, including all galaxies with $M_* > 10^{9}M_\odot/h$ and $M_* > 10^{10}M_\odot/h$, as well as splits of these samples into red/blue galaxies and highly/less star forming galaxies. These splits are shown in Fig.~\ref{fig:fig_sigma8part_scatter}. The red/blue split at $z=0$ is the same used by Ref.~\cite{2018MNRAS.475..676S} using the same simulation data (see their Eq.~(17)), but the remainder of the splits are simply guided by eye to give sufficiently different galaxy populations with which to test the forward modeling approach. These results were obtained with $\Lambda = 0.2h/{\rm Mpc}$ and all these galaxy samples satisfy the criterion $r(k) > 0.5$ for $k < \Lambda$ mentioned at the start of Sec.~\ref{sec:biasresults}.

For the samples selected just by the minimum stellar mass cut (black points in Fig.~\ref{fig:fig_sigma8part_Lambda_convergence_tng300_1_subhalo_hydro_stemass_splits_red}), the EFT likelihood recovers the expected value of $\sigma_8$ to within $1\%$ at all redshifts shown. As expected by the increased stochasticity with decreased galaxy number density, the performance degrades slightly when the color and ${\rm sSFR}$ cuts are made, but they remain within $5\%$ of the true value in most of the cases. The main exceptions to this are the red galaxies with $M_* > 10^{9}M_\odot/h$ at $z=0.5, 1$ and $M_* > 10^{10}M_\odot/h$ at $z=1$. Figure \ref{fig:fig_sigma8part_scatter} shows that these samples contain comparatively fewer objects compared to the others, and so it is unsurprising that the performance is somewhat poorer for these cases. The contribution from the missing fourth-order (and beyond) bias terms in our forward model is also more important for these more biased samples, which can also partly explain the observed shift in the inferred $\sigma_8$.

Recall, since we marginalize over all bias parameters up to third-order, which includes the linear LIMD parameter $b_1$, the extraction of the true value of $\sigma_8$ is thus achieved via nonlinear information that one would normally try to access with higher-order $N$-point correlation functions. In real-life applications, one must sample also the remainder of the cosmological parameters, as well as the initial conditions field $\delta_{m, \rm in}$, which should naturally reduce the constraining power on $\sigma_8$ (note however the very small volume $L_{\rm box} = 205 {\rm Mpc}/h$ of our samples compared to real-life surveys). The result shown in Fig.~\ref{fig:fig_sigma8part_Lambda_convergence_tng300_1_subhalo_hydro_stemass_splits_red} serves nonetheless as a strong validation test of the EFT likelihood of Eq.~(\ref{eq:eftlike}), and is a nice demonstration of the potential of the forward modeling approach to constrain cosmology using galaxy surveys. 

\section{Summary and conclusions}
\label{sec:summary}

Studying galaxy bias, the relation between the galaxy and underlying matter/energy distributions, is important not only to understand theoretically the connection between galaxy formation and the long-wavelength environment, but also because it can help to reduce the size of the parameter spaces explored in cosmological constraints using galaxy clustering data, and thus determine the cosmological parameters more precisely. In this paper we focused on the relations between the two second-order galaxy bias parameters $b_2$ and $b_{K^2}$, and the leading-order linear bias $b_1$; these relations appear, in particular, in analyses of the galaxy power spectrum starting at the 1-loop level and galaxy bispectrum starting at tree level. Past works on these parameters have focused on the case of halos in gravity-only simulations. Here, for the first time to the best of our knowledge, we took steps to investigate the bias relations for self-consistently simulated galaxies in hydrodynamical simulations. 

Concretely, we considered galaxies simulated with the \tng\ galaxy formation model (we used the TNG300-1 simulation with $L_{\rm box} = 205{\rm Mpc}/h$, $N_p = 2\times2500^3$), and estimated their bias parameters, for the first time also, using field-level forward models and the EFT likelihood formalism (cf.~Sec.~\ref{sec:method}). This formalism carries out the inference analysis directly at the galaxy density field level, which maximizes the utilization of the nonlinear information encoded in the galaxy distribution, compared to more standard approaches based on $N$-point correlation functions. One of our main goals was to contrast the shape of the $b_2(b_1)$ and $b_{K^2}(b_1)$ relations for objects selected by their total mass (which have been studied in past works) with the same relations for objects selected by a number of galaxy properties, in order to determine the extent to which these relations can be affected by galaxy physics and feedback effects. In addition to the galaxy bias parameters, in a separate part of our analysis, we have applied the EFT likelihood formalism to infer the value of $\sigma_8$ using the \tng\ galaxies as tracers.

Our main results can be summarized as follows:

\begin{itemize}

\item To validate the forward modeling and EFT likelihood approach to study galaxy bias, we applied it first to the case of halos from gravity-only simulations, and found the inferred values of $b_1$, $b_2$ and $b_{K^2}$ to be in good agreement with past works (cf.~Figs.~\ref{fig:fig_biaspart_Lambda_convergence_b2_both_tng300_1_tng800_1_group_dmo_totmass}, \ref{fig:fig_biaspart_Lambda_convergence_bK2_both_tng300_1_tng800_1_group_dmo_totmass}, \ref{fig:fig_biaspart_Lambda_convergence_0p2_b1_both_tng300_1_tng800_1_group_dmo_totmass}).

\item The $b_2(b_1)$ and $b_{K^2}(b_1)$ relations for galaxies selected by stellar mass, galaxy color, black hole accretion rate and specific star formation rate, are broadly the same as for objects selected by their total mass (cf.~grey vs.~colored points in Figs.~\ref{fig:fig_biaspart_Lambda_02_tng300_1_part1} and \ref{fig:fig_biaspart_Lambda_02_tng300_1_part2}). These relations for simulated galaxies are also in broad agreement with recent measurements of observed galaxy samples (cf.~Fig.~\ref{fig:obs}).

There were a few potentially interesting differences to the total mass selection results, e.g.~in the values of $b_2$ for stellar mass selected galaxies (cf.~lower center panel in Fig.~\ref{fig:fig_biaspart_Lambda_02_tng300_1_part1}) or the values of $b_{K^2}$ for color selected galaxies (cf.~upper right panel of Fig.~\ref{fig:fig_biaspart_Lambda_02_tng300_1_part2}), but a more in-depth study of these is needed. Using the halo model, we showed in Sec.~\ref{sec:hm} how differences like these could be physically linked to the way the relation between different galaxy properties and mass depends on large-scale overdensities or tidal fields.

\item Even after marginalizing over all galaxy bias parameters, the forward model and the EFT likelihood recover the true value of $\sigma_8$ to within $1\%$ using as tracers galaxies selected by a minimum stellar mass cut (cf.~Fig.~\ref{fig:fig_sigma8part_Lambda_convergence_tng300_1_subhalo_hydro_stemass_splits_red}). This precision degraded slightly to $\approx 5\%$ when these samples were subsequently divided into red/blue and highly/less star forming galaxies, which is still satisfactory given the lower number density after these splits are made and the small volume of the sample ($L_{\rm box} = 205{\rm Mpc}/h$).  

These results confirm that the EFT likelihood is able to yield unbiased cosmology constraints from nonlinear galaxy clustering even when selecting tracers (simulated galaxies) which are strongly influenced by baryonic effects such as gas cooling and stellar/black hole feedback.

\end{itemize}

The fact that the $b_2(b_1)$ and $b_{K^2}(b_1)$ relations of self-consistently simulated galaxies are not dramatically different from that of total mass selected objects can be encouraging as it suggests that priors based on the $b_2(b_1)$ and $b_{K^2}(b_1)$ relations shown in Figs.~\ref{fig:fig_biaspart_Lambda_02_tng300_1_part1} and \ref{fig:fig_biaspart_Lambda_02_tng300_1_part2} may be added to cosmological constraint analyses using galaxy clustering data to obtain tighter constraints without biasing them. These priors may take, for instance, the gravity-only result for total mass selection as the mean value, with some width informed by hydrodynamical simulations spanning a range of galaxy physics models and cosmologies (note that although the values of the bias parameters can depend on cosmology, the $b_2(b_1)$ and $b_{K^2}(b_1)$ relations should have a weaker dependence, as suggested by their weak redshift-dependence); the shaded map in Fig.~\ref{fig:obs} shows an example of such priors.

It is worth highlighting the efficiency of the forward modeling approach to study higher-order bias parameters such as $b_2$ and $b_{K^2}$. These two parameters can also be fitted for using the galaxy power spectrum and bispectrum, but accurate measurements of these statistics (especially the bispectrum) require large simulation volumes. The recent in-depth and comprehensive work of Ref.~\cite{2021arXiv210206902E} followed this approach using several realizations of gravity-only simulations with sizes ranging between $L_{\rm box} = 1000{\rm Mpc}/h$ and $L_{\rm box} = 2400{\rm Mpc}/h$, but these simulation volumes are currently numerically prohibitive with galaxy formation models like \tng. Methods based on separate universe simulations \cite{baldauf/etal:2015, lazeyras/etal, li/hu/takada:2016, 2018PhRvD..97l3526C, 2019MNRAS.488.2079B, 2020JCAP...12..013B, andreas, 2020arXiv201106584A, 2020MNRAS.496..483M, 2021MNRAS.503.1473S} do not strictly require large simulation volumes, but estimates of $b_1$ and $b_2$ still involve re-running the simulations for cosmological parameters that mimic different values of the long-wavelength overdensity. The same is true for separate universe simulations with tidal fields \cite{andreas, 2020arXiv201106584A, 2020MNRAS.496..483M, 2021MNRAS.503.1473S}, although these have not yet been performed with galaxy formation simulations. On the other hand, as we have seen in this paper, forward models are able to return satisfactory estimates of $b_2$ and $b_{K^2}$ using the relatively small-volume, high-resolution simulations of galaxy formation that already exist.

As future work, it would be interesting to follow up on the small differences of the galaxy $b_2(b_1)$ and $b_{K^2}(b_1)$ relations, compared to halos/subhalos, especially since as we discussed in Sec.~\ref{sec:hm}, they may encode information about the galaxy-environment connection on large scales. These future developments should include an extension of our analysis here to include beyond third-order bias operators, as well as an improved treatment of operator renormalization, to understand the origin of the small $\Lambda$-dependences observed in Fig.~\ref{fig:fig_biaspart_Lambda_convergence_b2_both_tng300_1_tng800_1_group_dmo_totmass} and discussed in Sec.~\ref{sec:lambda}. This would also allow us to begin studying the third-order bias parameters of simulated galaxies, which are relevant to cosmological constraint analyses with the galaxy power spectrum and bispectrum at the 1-loop level. Finally, it would be interesting to repeat the analysis here on other state-of-the-art galaxy formation models to broaden the range of astrophysical implementations explored; beyond galaxies as tracers, the EFT likelihood formalism can also be straightforwardly generalized to study the bias of the gas distribution, which is relevant to analyses of line-intensity mapping data.

\acknowledgments

We would like to thank Nico Hamaus and Volker Springel for useful comments, the \tng\ team for making their simulation data publicly available and Mikhail Ivanov for sharing the constraints on the bias parameters of the eBOSS ELG sample in advance. AB acknowledges support from the Excellence Cluster ORIGINS which is funded by the Deutsche Forschungsgemeinschaft (DFG, German Research Foundation) under Germany's Excellence Strategy - EXC-2094-390783311. TL is supported by the INFN INDARK grant. FS acknowledges support from the Starting Grant (ERC-2015-STG 678652) ``GrInflaGal'' of the European Research Council. The numerical analysis of the simulation data presented in this work was done on the Freya supercomputer at the Max Planck Computing and Data Facility (MPCDF) in Garching near Munich. 

\bibliographystyle{utphys}
\bibliography{REFS}

\providecommand{\href}[2]{#2}\begingroup\raggedright\begin{thebibliography}{10}

\bibitem{biasreview}
V.~{Desjacques}, D.~{Jeong}, and F.~{Schmidt}, ``{Large-scale galaxy bias},''
  \href{http://dx.doi.org/10.1016/j.physrep.2017.12.002}{{\em \physrep}
  {\bfseries 733} (Feb., 2018) 1--193},
  \href{http://arxiv.org/abs/1611.09787}{{\ttfamily arXiv:1611.09787}}.

\bibitem{baumann}
D.~{Baumann}, A.~{Nicolis}, L.~{Senatore}, and M.~{Zaldarriaga},
  ``{Cosmological non-linearities as an effective fluid},''
  \href{http://dx.doi.org/10.1088/1475-7516/2012/07/051}{{\em \jcap} {\bfseries
  7} (July, 2012) 051}, \href{http://arxiv.org/abs/1004.2488}{{\ttfamily
  arXiv:1004.2488}}.

\bibitem{2012JHEP...09..082C}
J.~J.~M. {Carrasco}, M.~P. {Hertzberg}, and L.~{Senatore}, ``{The effective
  field theory of cosmological large scale structures},''
  \href{http://dx.doi.org/10.1007/JHEP09(2012)082}{{\em Journal of High Energy
  Physics} {\bfseries 9} (Sept., 2012) 82},
  \href{http://arxiv.org/abs/1206.2926}{{\ttfamily arXiv:1206.2926
  [astro-ph.CO]}}.

\bibitem{Bernardeau/etal:2002}
F.~{Bernardeau}, S.~{Colombi}, E.~{Gazta{\~n}aga}, and R.~{Scoccimarro},
  ``{Large-scale structure of the Universe and cosmological perturbation
  theory},'' \href{http://dx.doi.org/10.1016/S0370-1573(02)00135-7}{{\em
  \physrep} {\bfseries 367} (Sept., 2002) 1--248},
  \href{http://arxiv.org/abs/arXiv:astro-ph/0112551}{{\ttfamily
  arXiv:astro-ph/0112551}}.

\bibitem{2016arXiv160200674B}
T.~{Baldauf}, M.~{Mirbabayi}, M.~{Simonovi{\'c}}, and M.~{Zaldarriaga}, ``{LSS
  constraints with controlled theoretical uncertainties},'' {\em arXiv
  e-prints} (Feb., 2016) arXiv:1602.00674,
  \href{http://arxiv.org/abs/1602.00674}{{\ttfamily arXiv:1602.00674
  [astro-ph.CO]}}.

\bibitem{2020PhRvD.102l3521W}
D.~{Wadekar}, M.~M. {Ivanov}, and R.~{Scoccimarro}, ``{Cosmological constraints
  from BOSS with analytic covariance matrices},''
  \href{http://dx.doi.org/10.1103/PhysRevD.102.123521}{{\em \prd} {\bfseries
  102} no.~12, (Dec., 2020) 123521},
  \href{http://arxiv.org/abs/2009.00622}{{\ttfamily arXiv:2009.00622
  [astro-ph.CO]}}.

\bibitem{fry/gaztanaga:1983}
J.~N. {Fry} and E.~{Gaztanaga}, ``{Biasing and hierarchical statistics in
  large-scale structure},'' \href{http://dx.doi.org/10.1086/173015}{{\em \apj}
  {\bfseries 413} (Aug., 1993) 447--452},
  \href{http://arxiv.org/abs/arXiv:astro-ph/9302009}{{\ttfamily
  arXiv:astro-ph/9302009}}.

\bibitem{2010ApJ...724..878T}
J.~L. {Tinker}, B.~E. {Robertson}, A.~V. {Kravtsov}, A.~{Klypin}, M.~S.
  {Warren}, G.~{Yepes}, and S.~{Gottl{\"o}ber}, ``{The Large-scale Bias of Dark
  Matter Halos: Numerical Calibration and Model Tests},''
  \href{http://dx.doi.org/10.1088/0004-637X/724/2/878}{{\em \apj} {\bfseries
  724} no.~2, (Dec, 2010) 878--886},
  \href{http://arxiv.org/abs/1001.3162}{{\ttfamily arXiv:1001.3162
  [astro-ph.CO]}}.

\bibitem{lazeyras/etal}
T.~{Lazeyras}, C.~{Wagner}, T.~{Baldauf}, and F.~{Schmidt}, ``{Precision
  measurement of the local bias of dark matter halos},''
  \href{http://dx.doi.org/10.1088/1475-7516/2016/02/018}{{\em \jcap} {\bfseries
  2} (Feb., 2016) 018}, \href{http://arxiv.org/abs/1511.01096}{{\ttfamily
  arXiv:1511.01096}}.

\bibitem{2017MNRAS.465.2225H}
K.~{Hoffmann}, J.~{Bel}, and E.~{Gazta{\~n}aga}, ``{Linear and non-linear bias:
  predictions versus measurements},''
  \href{http://dx.doi.org/10.1093/mnras/stw2876}{{\em \mnras} {\bfseries 465}
  no.~2, (Feb., 2017) 2225--2235},
  \href{http://arxiv.org/abs/1607.01024}{{\ttfamily arXiv:1607.01024
  [astro-ph.CO]}}.

\bibitem{mcdonald/roy:2009}
P.~{McDonald} and A.~{Roy}, ``{Clustering of dark matter tracers: generalizing
  bias for the coming era of precision LSS},''
  \href{http://dx.doi.org/10.1088/1475-7516/2009/08/020}{{\em \jcap} {\bfseries
  8} (Aug., 2009) 20}, \href{http://arxiv.org/abs/0902.0991}{{\ttfamily
  arXiv:0902.0991 [astro-ph.CO]}}.

\bibitem{sheth/chan/scoccimarro:2012}
R.~K. {Sheth}, K.~C. {Chan}, and R.~{Scoccimarro}, ``{Nonlocal Lagrangian
  bias},'' \href{http://dx.doi.org/10.1103/PhysRevD.87.083002}{{\em \prd}
  {\bfseries 87} no.~8, (Apr., 2013) 083002},
  \href{http://arxiv.org/abs/1207.7117}{{\ttfamily arXiv:1207.7117
  [astro-ph.CO]}}.

\bibitem{chan/scoccimarro/sheth:2012}
K.~C. {Chan}, R.~{Scoccimarro}, and R.~K. {Sheth}, ``{Gravity and large-scale
  nonlocal bias},'' \href{http://dx.doi.org/10.1103/PhysRevD.85.083509}{{\em
  \prd} {\bfseries 85} no.~8, (Apr., 2012) 083509},
  \href{http://arxiv.org/abs/1201.3614}{{\ttfamily arXiv:1201.3614
  [astro-ph.CO]}}.

\bibitem{baldauf/etal:2012}
T.~{Baldauf}, U.~{Seljak}, V.~{Desjacques}, and P.~{McDonald}, ``{Evidence for
  Quadratic Tidal Tensor Bias from the Halo Bispectrum},'' {\em ArXiv e-prints}
  (Jan., 2012) , \href{http://arxiv.org/abs/1201.4827}{{\ttfamily
  arXiv:1201.4827 [astro-ph.CO]}}.

\bibitem{MSZ}
M.~{Mirbabayi}, F.~{Schmidt}, and M.~{Zaldarriaga}, ``{Biased tracers and time
  evolution},'' \href{http://dx.doi.org/10.1088/1475-7516/2015/07/030}{{\em
  \jcap} {\bfseries 7} (July, 2015) 30},
  \href{http://arxiv.org/abs/1412.5169}{{\ttfamily arXiv:1412.5169}}.

\bibitem{2018JCAP...09..008L}
T.~{Lazeyras} and F.~{Schmidt}, ``{Beyond LIMD bias: a measurement of the
  complete set of third-order halo bias parameters},''
  \href{http://dx.doi.org/10.1088/1475-7516/2018/09/008}{{\em \jcap} {\bfseries
  2018} no.~9, (Sept., 2018) 008},
  \href{http://arxiv.org/abs/1712.07531}{{\ttfamily arXiv:1712.07531
  [astro-ph.CO]}}.

\bibitem{saito/etal:14}
S.~{Saito}, T.~{Baldauf}, Z.~{Vlah}, U.~{Seljak}, T.~{Okumura}, and
  P.~{McDonald}, ``{Understanding higher-order nonlocal halo bias at large
  scales by combining the power spectrum with the bispectrum},''
  \href{http://dx.doi.org/10.1103/PhysRevD.90.123522}{{\em \prd} {\bfseries 90}
  no.~12, (Dec., 2014) 123522},
  \href{http://arxiv.org/abs/1405.1447}{{\ttfamily arXiv:1405.1447}}.

\bibitem{2018JCAP...07..029A}
M.~M. {Abidi} and T.~{Baldauf}, ``{Cubic halo bias in Eulerian and Lagrangian
  space},'' \href{http://dx.doi.org/10.1088/1475-7516/2018/07/029}{{\em \jcap}
  {\bfseries 2018} no.~7, (July, 2018) 029},
  \href{http://arxiv.org/abs/1802.07622}{{\ttfamily arXiv:1802.07622
  [astro-ph.CO]}}.

\bibitem{2020PhRvD.102j3530E}
A.~{Eggemeier}, R.~{Scoccimarro}, M.~{Crocce}, A.~{Pezzotta}, and A.~G.
  {S{\'a}nchez}, ``{Testing one-loop galaxy bias: Power spectrum},''
  \href{http://dx.doi.org/10.1103/PhysRevD.102.103530}{{\em \prd} {\bfseries
  102} no.~10, (Nov., 2020) 103530},
  \href{http://arxiv.org/abs/2006.09729}{{\ttfamily arXiv:2006.09729
  [astro-ph.CO]}}.

\bibitem{2021arXiv210206902E}
A.~{Eggemeier}, R.~{Scoccimarro}, R.~E. {Smith}, M.~{Crocce}, A.~{Pezzotta},
  and A.~G. {S{\'a}nchez}, ``{Testing one-loop galaxy bias: joint analysis of
  power spectrum and bispectrum},'' {\em arXiv e-prints} (Feb., 2021)
  arXiv:2102.06902, \href{http://arxiv.org/abs/2102.06902}{{\ttfamily
  arXiv:2102.06902 [astro-ph.CO]}}.

\bibitem{2019JCAP...11..041L}
T.~{Lazeyras} and F.~{Schmidt}, ``{A robust measurement of the first
  higher-derivative bias of dark matter halos},''
  \href{http://dx.doi.org/10.1088/1475-7516/2019/11/041}{{\em \jcap} {\bfseries
  2019} no.~11, (Nov., 2019) 041},
  \href{http://arxiv.org/abs/1904.11294}{{\ttfamily arXiv:1904.11294
  [astro-ph.CO]}}.

\bibitem{dalal/etal:2008}
N.~{Dalal}, O.~{Dor{\'e}}, D.~{Huterer}, and A.~{Shirokov}, ``{Imprints of
  primordial non-Gaussianities on large-scale structure: Scale-dependent bias
  and abundance of virialized objects},''
  \href{http://dx.doi.org/10.1103/PhysRevD.77.123514}{{\em \prd} {\bfseries 77}
  no.~12, (June, 2008) 123514--+},
  \href{http://arxiv.org/abs/0710.4560}{{\ttfamily arXiv:0710.4560}}.

\bibitem{giannantonio/porciani:2010}
T.~{Giannantonio} and C.~{Porciani}, ``{Structure formation from non-Gaussian
  initial conditions: Multivariate biasing, statistics, and comparison with
  N-body simulations},''
  \href{http://dx.doi.org/10.1103/PhysRevD.81.063530}{{\em \prd} {\bfseries 81}
  no.~6, (Mar., 2010) 063530--+},
  \href{http://arxiv.org/abs/0911.0017}{{\ttfamily arXiv:0911.0017}}.

\bibitem{assassi/baumann/schmidt}
V.~{Assassi}, D.~{Baumann}, and F.~{Schmidt}, ``{Galaxy bias and primordial
  non-Gaussianity},''
  \href{http://dx.doi.org/10.1088/1475-7516/2015/12/043}{{\em \jcap} {\bfseries
  12} (Dec., 2015) 043}, \href{http://arxiv.org/abs/1510.03723}{{\ttfamily
  arXiv:1510.03723}}.

\bibitem{2020JCAP...12..013B}
A.~{Barreira}, G.~{Cabass}, F.~{Schmidt}, A.~{Pillepich}, and D.~{Nelson},
  ``{Galaxy bias and primordial non-Gaussianity: insights from galaxy formation
  simulations with IllustrisTNG},''
  \href{http://dx.doi.org/10.1088/1475-7516/2020/12/013}{{\em \jcap} {\bfseries
  2020} no.~12, (Dec., 2020) 013},
  \href{http://arxiv.org/abs/2006.09368}{{\ttfamily arXiv:2006.09368
  [astro-ph.CO]}}.

\bibitem{2020arXiv201014523M}
A.~{Moradinezhad Dizgah}, M.~{Biagetti}, E.~{Sefusatti}, V.~{Desjacques}, and
  J.~{Nore{\~n}a}, ``{Primordial Non-Gaussianity from Biased Tracers:
  Likelihood Analysis of Real-Space Power Spectrum and Bispectrum},'' {\em
  arXiv e-prints} (Oct., 2020) arXiv:2010.14523,
  \href{http://arxiv.org/abs/2010.14523}{{\ttfamily arXiv:2010.14523
  [astro-ph.CO]}}.

\bibitem{2020JCAP...12..031B}
A.~{Barreira}, ``{On the impact of galaxy bias uncertainties on primordial
  non-Gaussianity constraints},''
  \href{http://dx.doi.org/10.1088/1475-7516/2020/12/031}{{\em \jcap} {\bfseries
  2020} no.~12, (Dec., 2020) 031},
  \href{http://arxiv.org/abs/2009.06622}{{\ttfamily arXiv:2009.06622
  [astro-ph.CO]}}.

\bibitem{tseliakhovich/hirata:2010}
D.~{Tseliakhovich} and C.~{Hirata}, ``{Relative velocity of dark matter and
  baryonic fluids and the formation of the first structures},''
  \href{http://dx.doi.org/10.1103/PhysRevD.82.083520}{{\em \prd} {\bfseries 82}
  no.~8, (Oct., 2010) 083520}, \href{http://arxiv.org/abs/1005.2416}{{\ttfamily
  arXiv:1005.2416 [astro-ph.CO]}}.

\bibitem{2016PhRvD..94f3508S}
F.~{Schmidt}, ``{Effect of relative velocity and density perturbations between
  baryons and dark matter on the clustering of galaxies},''
  \href{http://dx.doi.org/10.1103/PhysRevD.94.063508}{{\em \prd} {\bfseries 94}
  no.~6, (Sept., 2016) 063508},
  \href{http://arxiv.org/abs/1602.09059}{{\ttfamily arXiv:1602.09059}}.

\bibitem{2020JCAP...02..005B}
A.~{Barreira}, G.~{Cabass}, D.~{Nelson}, and F.~{Schmidt}, ``{Baryon-CDM
  isocurvature galaxy bias with IllustrisTNG},''
  \href{http://dx.doi.org/10.1088/1475-7516/2020/02/005}{{\em \jcap} {\bfseries
  2020} no.~2, (Feb, 2020) 005},
  \href{http://arxiv.org/abs/1907.04317}{{\ttfamily arXiv:1907.04317
  [astro-ph.CO]}}.

\bibitem{2021JCAP...03..023K}
H.~{Khoraminezhad}, T.~{Lazeyras}, R.~E. {Angulo}, O.~{Hahn}, and M.~{Viel},
  ``{Quantifying the impact of baryon-CDM perturbations on halo clustering and
  baryon fraction},''
  \href{http://dx.doi.org/10.1088/1475-7516/2021/03/023}{{\em \jcap} {\bfseries
  2021} no.~3, (Mar., 2021) 023},
  \href{http://arxiv.org/abs/2011.01037}{{\ttfamily arXiv:2011.01037
  [astro-ph.CO]}}.

\bibitem{2017PhRvD..96h3533S}
F.~{Schmidt} and F.~{Beutler}, ``{Imprints of reionization in galaxy
  clustering},'' \href{http://dx.doi.org/10.1103/PhysRevD.96.083533}{{\em \prd}
  {\bfseries 96} no.~8, (Oct, 2017) 083533},
  \href{http://arxiv.org/abs/1705.07843}{{\ttfamily arXiv:1705.07843
  [astro-ph.CO]}}.

\bibitem{2019JCAP...05..031C}
G.~{Cabass} and F.~{Schmidt}, ``{A new scale in the bias expansion},''
  \href{http://dx.doi.org/10.1088/1475-7516/2019/05/031}{{\em Journal of
  Cosmology and Astro-Particle Physics} {\bfseries 2019} no.~5, (May, 2019)
  031}, \href{http://arxiv.org/abs/1812.02731}{{\ttfamily arXiv:1812.02731
  [astro-ph.CO]}}.

\bibitem{2020JCAP...07..049B}
A.~{Barreira}, G.~{Cabass}, K.~D. {Lozanov}, and F.~{Schmidt}, ``{Compensated
  isocurvature perturbations in the galaxy power spectrum},''
  \href{http://dx.doi.org/10.1088/1475-7516/2020/07/049}{{\em \jcap} {\bfseries
  2020} no.~7, (July, 2020) 049},
  \href{http://arxiv.org/abs/2002.12931}{{\ttfamily arXiv:2002.12931
  [astro-ph.CO]}}.

\bibitem{2014MNRAS.444.1518V}
M.~{Vogelsberger}, S.~{Genel}, V.~{Springel}, P.~{Torrey}, D.~{Sijacki},
  D.~{Xu}, G.~{Snyder}, D.~{Nelson}, and L.~{Hernquist}, ``{Introducing the
  Illustris Project: simulating the coevolution of dark and visible matter in
  the Universe},'' \href{http://dx.doi.org/10.1093/mnras/stu1536}{{\em \mnras}
  {\bfseries 444} (Oct, 2014) 1518--1547},
  \href{http://arxiv.org/abs/1405.2921}{{\ttfamily arXiv:1405.2921
  [astro-ph.CO]}}.

\bibitem{2015MNRAS.446..521S}
J.~{Schaye}, R.~A. {Crain}, R.~G. {Bower}, M.~{Furlong}, M.~{Schaller},
  T.~{Theuns}, C.~{Dalla Vecchia}, C.~S. {Frenk}, I.~G. {McCarthy}, J.~C.
  {Helly}, A.~{Jenkins}, Y.~M. {Rosas-Guevara}, S.~D.~M. {White}, M.~{Baes},
  C.~M. {Booth}, P.~{Camps}, J.~F. {Navarro}, Y.~{Qu}, A.~{Rahmati},
  T.~{Sawala}, P.~A. {Thomas}, and J.~{Trayford}, ``{The EAGLE project:
  simulating the evolution and assembly of galaxies and their environments},''
  \href{http://dx.doi.org/10.1093/mnras/stu2058}{{\em \mnras} {\bfseries 446}
  no.~1, (Jan., 2015) 521--554},
  \href{http://arxiv.org/abs/1407.7040}{{\ttfamily arXiv:1407.7040
  [astro-ph.GA]}}.

\bibitem{2017arXiv170609899T}
{The EAGLE team}, ``{The EAGLE simulations of galaxy formation: Public release
  of particle data},'' {\em arXiv e-prints} (June, 2017) arXiv:1706.09899,
  \href{http://arxiv.org/abs/1706.09899}{{\ttfamily arXiv:1706.09899
  [astro-ph.GA]}}.

\bibitem{2014MNRAS.442.2304H}
M.~{Hirschmann}, K.~{Dolag}, A.~{Saro}, L.~{Bachmann}, S.~{Borgani}, and
  A.~{Burkert}, ``{Cosmological simulations of black hole growth: AGN
  luminosities and downsizing},''
  \href{http://dx.doi.org/10.1093/mnras/stu1023}{{\em \mnras} {\bfseries 442}
  no.~3, (Aug., 2014) 2304--2324},
  \href{http://arxiv.org/abs/1308.0333}{{\ttfamily arXiv:1308.0333
  [astro-ph.CO]}}.

\bibitem{2017MNRAS.465.2936M}
I.~G. {McCarthy}, J.~{Schaye}, S.~{Bird}, and A.~M.~C. {Le Brun}, ``{The
  BAHAMAS project: calibrated hydrodynamical simulations for large-scale
  structure cosmology},'' \href{http://dx.doi.org/10.1093/mnras/stw2792}{{\em
  \mnras} {\bfseries 465} no.~3, (Mar., 2017) 2936--2965},
  \href{http://arxiv.org/abs/1603.02702}{{\ttfamily arXiv:1603.02702
  [astro-ph.CO]}}.

\bibitem{2014MNRAS.444.1453D}
Y.~{Dubois}, C.~{Pichon}, C.~{Welker}, D.~{Le Borgne}, J.~{Devriendt},
  C.~{Laigle}, S.~{Codis}, D.~{Pogosyan}, S.~{Arnouts}, K.~{Benabed},
  E.~{Bertin}, J.~{Blaizot}, F.~{Bouchet}, J.~F. {Cardoso}, S.~{Colombi},
  V.~{de Lapparent}, V.~{Desjacques}, R.~{Gavazzi}, S.~{Kassin}, T.~{Kimm},
  H.~{McCracken}, B.~{Milliard}, S.~{Peirani}, S.~{Prunet}, S.~{Rouberol},
  J.~{Silk}, A.~{Slyz}, T.~{Sousbie}, R.~{Teyssier}, L.~{Tresse}, M.~{Treyer},
  D.~{Vibert}, and M.~{Volonteri}, ``{Dancing in the dark: galactic properties
  trace spin swings along the cosmic web},''
  \href{http://dx.doi.org/10.1093/mnras/stu1227}{{\em \mnras} {\bfseries 444}
  no.~2, (Oct., 2014) 1453--1468},
  \href{http://arxiv.org/abs/1402.1165}{{\ttfamily arXiv:1402.1165
  [astro-ph.CO]}}.

\bibitem{Pillepich:2017jle}
A.~Pillepich {\em et~al.}, ``{Simulating Galaxy Formation with the IllustrisTNG
  Model},'' \href{http://dx.doi.org/10.1093/mnras/stx2656}{{\em Mon. Not. Roy.
  Astron. Soc.} {\bfseries 473} no.~3, (2018) 4077--4106},
\href{http://arxiv.org/abs/1703.02970}{{\ttfamily arXiv:1703.02970
  [astro-ph.GA]}}.

\bibitem{2017MNRAS.465.3291W}
R.~{Weinberger}, V.~{Springel}, L.~{Hernquist}, A.~{Pillepich}, F.~{Marinacci},
  R.~{Pakmor}, D.~{Nelson}, S.~{Genel}, M.~{Vogelsberger}, J.~{Naiman}, and
  P.~{Torrey}, ``{Simulating galaxy formation with black hole driven thermal
  and kinetic feedback},'' \href{http://dx.doi.org/10.1093/mnras/stw2944}{{\em
  \mnras} {\bfseries 465} (Mar, 2017) 3291--3308},
  \href{http://arxiv.org/abs/1607.03486}{{\ttfamily arXiv:1607.03486
  [astro-ph.GA]}}.

\bibitem{Nelson:2018uso}
D.~Nelson {\em et~al.}, ``{The IllustrisTNG Simulations: Public Data
  Release},'' {\em arXiv:1812.05609} (2018) ,
\href{http://arxiv.org/abs/1812.05609}{{\ttfamily arXiv:1812.05609
  [astro-ph.GA]}}.

\bibitem{2010MNRAS.409..355J}
J.~{Jasche}, F.~S. {Kitaura}, C.~{Li}, and T.~A. {En{\ss}lin}, ``{Bayesian
  non-linear large-scale structure inference of the Sloan Digital Sky Survey
  Data Release 7},''
  \href{http://dx.doi.org/10.1111/j.1365-2966.2010.17313.x}{{\em \mnras}
  {\bfseries 409} no.~1, (Nov., 2010) 355--370},
  \href{http://arxiv.org/abs/0911.2498}{{\ttfamily arXiv:0911.2498
  [astro-ph.CO]}}.

\bibitem{2013MNRAS.432..894J}
J.~{Jasche} and B.~D. {Wandelt}, ``{Bayesian physical reconstruction of initial
  conditions from large-scale structure surveys},''
  \href{http://dx.doi.org/10.1093/mnras/stt449}{{\em \mnras} {\bfseries 432}
  no.~2, (June, 2013) 894--913},
  \href{http://arxiv.org/abs/1203.3639}{{\ttfamily arXiv:1203.3639
  [astro-ph.CO]}}.

\bibitem{2014ApJ...794...94W}
H.~{Wang}, H.~J. {Mo}, X.~{Yang}, Y.~P. {Jing}, and W.~P. {Lin},
  ``{ELUCID{\textemdash}Exploring the Local Universe with the Reconstructed
  Initial Density Field. I. Hamiltonian Markov Chain Monte Carlo Method with
  Particle Mesh Dynamics},''
  \href{http://dx.doi.org/10.1088/0004-637X/794/1/94}{{\em \apj} {\bfseries
  794} no.~1, (Oct., 2014) 94},
  \href{http://arxiv.org/abs/1407.3451}{{\ttfamily arXiv:1407.3451
  [astro-ph.CO]}}.

\bibitem{2015MNRAS.446.4250A}
M.~{Ata}, F.-S. {Kitaura}, and V.~{M{\"u}ller}, ``{Bayesian inference of cosmic
  density fields from non-linear, scale-dependent, and stochastic biased
  tracers},'' \href{http://dx.doi.org/10.1093/mnras/stu2347}{{\em \mnras}
  {\bfseries 446} no.~4, (Feb., 2015) 4250--4259},
  \href{http://arxiv.org/abs/1408.2566}{{\ttfamily arXiv:1408.2566
  [astro-ph.CO]}}.

\bibitem{2016ApJ...831..164W}
H.~{Wang}, H.~J. {Mo}, X.~{Yang}, Y.~{Zhang}, J.~{Shi}, Y.~P. {Jing}, C.~{Liu},
  S.~{Li}, X.~{Kang}, and Y.~{Gao}, ``{ELUCID - Exploring the Local Universe
  with ReConstructed Initial Density Field III: Constrained Simulation in the
  SDSS Volume},'' \href{http://dx.doi.org/10.3847/0004-637X/831/2/164}{{\em
  \apj} {\bfseries 831} no.~2, (Nov., 2016) 164},
  \href{http://arxiv.org/abs/1608.01763}{{\ttfamily arXiv:1608.01763
  [astro-ph.CO]}}.

\bibitem{2017JCAP...12..009S}
U.~{Seljak}, G.~{Aslanyan}, Y.~{Feng}, and C.~{Modi}, ``{Towards optimal
  extraction of cosmological information from nonlinear data},''
  \href{http://dx.doi.org/10.1088/1475-7516/2017/12/009}{{\em \jcap} {\bfseries
  2017} no.~12, (Dec., 2017) 009},
  \href{http://arxiv.org/abs/1706.06645}{{\ttfamily arXiv:1706.06645
  [astro-ph.CO]}}.

\bibitem{2019PhRvD.100d3514S}
M.~{Schmittfull}, M.~{Simonovi{\'c}}, V.~{Assassi}, and M.~{Zaldarriaga},
  ``{Modeling biased tracers at the field level},''
  \href{http://dx.doi.org/10.1103/PhysRevD.100.043514}{{\em \prd} {\bfseries
  100} no.~4, (Aug., 2019) 043514},
  \href{http://arxiv.org/abs/1811.10640}{{\ttfamily arXiv:1811.10640
  [astro-ph.CO]}}.

\bibitem{2019JCAP...11..023M}
C.~{Modi}, M.~{White}, A.~{Slosar}, and E.~{Castorina}, ``{Reconstructing
  large-scale structure with neutral hydrogen surveys},''
  \href{http://dx.doi.org/10.1088/1475-7516/2019/11/023}{{\em \jcap} {\bfseries
  2019} no.~11, (Nov., 2019) 023},
  \href{http://arxiv.org/abs/1907.02330}{{\ttfamily arXiv:1907.02330
  [astro-ph.CO]}}.

\bibitem{2020arXiv201203334S}
M.~{Schmittfull}, M.~{Simonovi{\'c}}, M.~M. {Ivanov}, O.~H.~E. {Philcox}, and
  M.~{Zaldarriaga}, ``{Modeling Galaxies in Redshift Space at the Field
  Level},'' {\em arXiv e-prints} (Dec., 2020) arXiv:2012.03334,
  \href{http://arxiv.org/abs/2012.03334}{{\ttfamily arXiv:2012.03334
  [astro-ph.CO]}}.

\bibitem{2019A&A...625A..64J}
J.~{Jasche} and G.~{Lavaux}, ``{Physical Bayesian modelling of the non-linear
  matter distribution: New insights into the nearby universe},''
  \href{http://dx.doi.org/10.1051/0004-6361/201833710}{{\em \aap} {\bfseries
  625} (May, 2019) A64}, \href{http://arxiv.org/abs/1806.11117}{{\ttfamily
  arXiv:1806.11117 [astro-ph.CO]}}.

\bibitem{2019arXiv190906396L}
G.~{Lavaux}, J.~{Jasche}, and F.~{Leclercq}, ``{Systematic-free inference of
  the cosmic matter density field from SDSS3-BOSS data},'' {\em arXiv e-prints}
  (Sept., 2019) arXiv:1909.06396,
  \href{http://arxiv.org/abs/1909.06396}{{\ttfamily arXiv:1909.06396
  [astro-ph.CO]}}.

\bibitem{2019JCAP...01..042S}
F.~{Schmidt}, F.~{Elsner}, J.~{Jasche}, N.~M. {Nguyen}, and G.~{Lavaux}, ``{A
  rigorous EFT-based forward model for large-scale structure},''
  \href{http://dx.doi.org/10.1088/1475-7516/2019/01/042}{{\em \jcap} {\bfseries
  2019} no.~1, (Jan., 2019) 042},
  \href{http://arxiv.org/abs/1808.02002}{{\ttfamily arXiv:1808.02002
  [astro-ph.CO]}}.

\bibitem{2020JCAP...01..029E}
F.~{Elsner}, F.~{Schmidt}, J.~{Jasche}, G.~{Lavaux}, and N.-M. {Nguyen},
  ``{Cosmology inference from a biased density field using the EFT-based
  likelihood},'' \href{http://dx.doi.org/10.1088/1475-7516/2020/01/029}{{\em
  \jcap} {\bfseries 2020} no.~1, (Jan., 2020) 029},
  \href{http://arxiv.org/abs/1906.07143}{{\ttfamily arXiv:1906.07143
  [astro-ph.CO]}}.

\bibitem{2020JCAP...11..008S}
F.~{Schmidt}, G.~{Cabass}, J.~{Jasche}, and G.~{Lavaux}, ``{Unbiased cosmology
  inference from biased tracers using the EFT likelihood},''
  \href{http://dx.doi.org/10.1088/1475-7516/2020/11/008}{{\em \jcap} {\bfseries
  2020} no.~11, (Nov., 2020) 008},
  \href{http://arxiv.org/abs/2004.06707}{{\ttfamily arXiv:2004.06707
  [astro-ph.CO]}}.

\bibitem{2020arXiv200914176S}
F.~{Schmidt}, ``{Sigma-eight at the percent level: the EFT likelihood in real
  space},'' \href{http://dx.doi.org/10.1088/1475-7516/2021/04/032}{{\em \jcap}
  {\bfseries 2021} no.~4, (Apr., 2021) 032},
  \href{http://arxiv.org/abs/2009.14176}{{\ttfamily arXiv:2009.14176
  [astro-ph.CO]}}.

\bibitem{2021JCAP...03..058N}
N.-M. {Nguyen}, F.~{Schmidt}, G.~{Lavaux}, and J.~{Jasche}, ``{Impacts of the
  physical data model on the forward inference of initial conditions from
  biased tracers},''
  \href{http://dx.doi.org/10.1088/1475-7516/2021/03/058}{{\em \jcap} {\bfseries
  2021} no.~3, (Mar., 2021) 058},
  \href{http://arxiv.org/abs/2011.06587}{{\ttfamily arXiv:2011.06587
  [astro-ph.CO]}}.

\bibitem{2020JCAP...04..042C}
G.~{Cabass} and F.~{Schmidt}, ``{The EFT likelihood for large-scale
  structure},'' \href{http://dx.doi.org/10.1088/1475-7516/2020/04/042}{{\em
  \jcap} {\bfseries 2020} no.~4, (Apr., 2020) 042},
  \href{http://arxiv.org/abs/1909.04022}{{\ttfamily arXiv:1909.04022
  [astro-ph.CO]}}.

\bibitem{2020JCAP...07..051C}
G.~{Cabass} and F.~{Schmidt}, ``{The likelihood for LSS: stochasticity of bias
  coefficients at all orders},''
  \href{http://dx.doi.org/10.1088/1475-7516/2020/07/051}{{\em \jcap} {\bfseries
  2020} no.~7, (July, 2020) 051},
  \href{http://arxiv.org/abs/2004.00617}{{\ttfamily arXiv:2004.00617
  [astro-ph.CO]}}.

\bibitem{Lazeyras:2021dar}
T.~Lazeyras, A.~Barreira, and F.~Schmidt, ``{Assembly bias in quadratic bias
  parameters of dark matter halos from forward modeling},''
  \href{http://arxiv.org/abs/2106.14713}{{\ttfamily arXiv:2106.14713
  [astro-ph.CO]}}.

\bibitem{2020arXiv201209837S}
F.~{Schmidt}, ``{An n-th order Lagrangian forward model for large-scale
  structure},'' \href{http://dx.doi.org/10.1088/1475-7516/2021/04/033}{{\em
  \jcap} {\bfseries 2021} no.~4, (Apr., 2021) 033},
  \href{http://arxiv.org/abs/2012.09837}{{\ttfamily arXiv:2012.09837
  [astro-ph.CO]}}.

\bibitem{hamaus/etal:2010}
N.~{Hamaus}, U.~{Seljak}, V.~{Desjacques}, R.~E. {Smith}, and T.~{Baldauf},
  ``{Minimizing the stochasticity of halos in large-scale structure surveys},''
  \href{http://dx.doi.org/10.1103/PhysRevD.82.043515}{{\em \prd} {\bfseries 82}
  no.~4, (Aug., 2010) 043515}, \href{http://arxiv.org/abs/1004.5377}{{\ttfamily
  arXiv:1004.5377 [astro-ph.CO]}}.

\bibitem{2018MNRAS.480.5113M}
F.~{Marinacci}, M.~{Vogelsberger}, R.~{Pakmor}, P.~{Torrey}, V.~{Springel},
  L.~{Hernquist}, D.~{Nelson}, R.~{Weinberger}, A.~{Pillepich}, J.~{Naiman},
  and S.~{Genel}, ``{First results from the IllustrisTNG simulations: radio
  haloes and magnetic fields},''
  \href{http://dx.doi.org/10.1093/mnras/sty2206}{{\em \mnras} {\bfseries 480}
  (Nov, 2018) 5113--5139}, \href{http://arxiv.org/abs/1707.03396}{{\ttfamily
  arXiv:1707.03396 [astro-ph.CO]}}.

\bibitem{Pillepich:2017fcc}
A.~Pillepich {\em et~al.}, ``{First results from the IllustrisTNG simulations:
  the stellar mass content of groups and clusters of galaxies},''
  \href{http://dx.doi.org/10.1093/mnras/stx3112}{{\em Mon. Not. Roy. Astron.
  Soc.} {\bfseries 475} (2018) 648},
\href{http://arxiv.org/abs/1707.03406}{{\ttfamily arXiv:1707.03406
  [astro-ph.GA]}}.

\bibitem{2018MNRAS.477.1206N}
J.~P. {Naiman}, A.~{Pillepich}, V.~{Springel}, E.~{Ramirez-Ruiz}, P.~{Torrey},
  M.~{Vogelsberger}, R.~{Pakmor}, D.~{Nelson}, F.~{Marinacci}, L.~{Hernquist},
  R.~{Weinberger}, and S.~{Genel}, ``{First results from the IllustrisTNG
  simulations: a tale of two elements - chemical evolution of magnesium and
  europium},'' \href{http://dx.doi.org/10.1093/mnras/sty618}{{\em \mnras}
  {\bfseries 477} (Jun, 2018) 1206--1224},
  \href{http://arxiv.org/abs/1707.03401}{{\ttfamily arXiv:1707.03401
  [astro-ph.GA]}}.

\bibitem{2018MNRAS.475..676S}
V.~{Springel}, R.~{Pakmor}, A.~{Pillepich}, R.~{Weinberger}, D.~{Nelson},
  L.~{Hernquist}, M.~{Vogelsberger}, S.~{Genel}, P.~{Torrey}, F.~{Marinacci},
  and J.~{Naiman}, ``{First results from the IllustrisTNG simulations: matter
  and galaxy clustering},'' \href{http://dx.doi.org/10.1093/mnras/stx3304}{{\em
  \mnras} {\bfseries 475} (Mar, 2018) 676--698},
  \href{http://arxiv.org/abs/1707.03397}{{\ttfamily arXiv:1707.03397
  [astro-ph.GA]}}.

\bibitem{Nelson:2017cxy}
D.~Nelson {\em et~al.}, ``{First results from the IllustrisTNG simulations: the
  galaxy color bimodality},''
  \href{http://dx.doi.org/10.1093/mnras/stx3040}{{\em Mon. Not. Roy. Astron.
  Soc.} {\bfseries 475} (2018) 624},
\href{http://arxiv.org/abs/1707.03395}{{\ttfamily arXiv:1707.03395
  [astro-ph.GA]}}.

\bibitem{2019MNRAS.490.3234N}
D.~{Nelson}, A.~{Pillepich}, V.~{Springel}, R.~{Pakmor}, R.~{Weinberger},
  S.~{Genel}, P.~{Torrey}, M.~{Vogelsberger}, F.~{Marinacci}, and
  L.~{Hernquist}, ``{First results from the TNG50 simulation: galactic outflows
  driven by supernovae and black hole feedback},''
  \href{http://dx.doi.org/10.1093/mnras/stz2306}{{\em \mnras} {\bfseries 490}
  no.~3, (Dec., 2019) 3234--3261},
  \href{http://arxiv.org/abs/1902.05554}{{\ttfamily arXiv:1902.05554
  [astro-ph.GA]}}.

\bibitem{2019MNRAS.490.3196P}
A.~{Pillepich}, D.~{Nelson}, V.~{Springel}, R.~{Pakmor}, P.~{Torrey},
  R.~{Weinberger}, M.~{Vogelsberger}, F.~{Marinacci}, S.~{Genel}, A.~{van der
  Wel}, and L.~{Hernquist}, ``{First results from the TNG50 simulation: the
  evolution of stellar and gaseous discs across cosmic time},''
  \href{http://dx.doi.org/10.1093/mnras/stz2338}{{\em \mnras} {\bfseries 490}
  no.~3, (Dec., 2019) 3196--3233},
  \href{http://arxiv.org/abs/1902.05553}{{\ttfamily arXiv:1902.05553
  [astro-ph.GA]}}.

\bibitem{2010MNRAS.401..791S}
V.~{Springel}, ``{E pur si muove: Galilean-invariant cosmological
  hydrodynamical simulations on a moving mesh},''
  \href{http://dx.doi.org/10.1111/j.1365-2966.2009.15715.x}{{\em \mnras}
  {\bfseries 401} (Jan, 2010) 791--851},
  \href{http://arxiv.org/abs/0901.4107}{{\ttfamily arXiv:0901.4107
  [astro-ph.CO]}}.

\bibitem{2016MNRAS.455.1134P}
R.~{Pakmor}, V.~{Springel}, A.~{Bauer}, P.~{Mocz}, D.~J. {Munoz}, S.~T.
  {Ohlmann}, K.~{Schaal}, and C.~{Zhu}, ``{Improving the convergence properties
  of the moving-mesh code AREPO},''
  \href{http://dx.doi.org/10.1093/mnras/stv2380}{{\em \mnras} {\bfseries 455}
  (Jan, 2016) 1134--1143}, \href{http://arxiv.org/abs/1503.00562}{{\ttfamily
  arXiv:1503.00562 [astro-ph.GA]}}.

\bibitem{2001MNRAS.328..726S}
V.~{Springel}, S.~D.~M. {White}, G.~{Tormen}, and G.~{Kauffmann}, ``{Populating
  a cluster of galaxies - I. Results at [formmu2]z=0},''
  \href{http://dx.doi.org/10.1046/j.1365-8711.2001.04912.x}{{\em \mnras}
  {\bfseries 328} (Dec., 2001) 726--750},
  \href{http://arxiv.org/abs/astro-ph/0012055}{{\ttfamily astro-ph/0012055}}.

\bibitem{2021JCAP...01..067C}
G.~{Cabass}, ``{The EFT likelihood for large-scale structure in redshift
  space},'' \href{http://dx.doi.org/10.1088/1475-7516/2021/01/067}{{\em \jcap}
  {\bfseries 2021} no.~1, (Jan., 2021) 067},
  \href{http://arxiv.org/abs/2007.14988}{{\ttfamily arXiv:2007.14988
  [astro-ph.CO]}}.

\bibitem{2020MNRAS.496.1182M}
A.~D. {Montero-Dorta}, M.~C. {Artale}, L.~R. {Abramo}, B.~{Tucci},
  N.~{Padilla}, G.~{Sato-Polito}, I.~{Lacerna}, F.~{Rodriguez}, and R.~E.
  {Angulo}, ``{The manifestation of secondary bias on the galaxy population
  from IllustrisTNG300},'' \href{http://dx.doi.org/10.1093/mnras/staa1624}{{\em
  \mnras} {\bfseries 496} no.~2, (Aug., 2020) 1182--1196},
  \href{http://arxiv.org/abs/2001.01739}{{\ttfamily arXiv:2001.01739
  [astro-ph.GA]}}.

\bibitem{2014MNRAS.442.2131A}
R.~E. {Angulo}, S.~D.~M. {White}, V.~{Springel}, and B.~{Henriques}, ``{Galaxy
  formation on the largest scales: the impact of astrophysics on the baryonic
  acoustic oscillation peak},''
  \href{http://dx.doi.org/10.1093/mnras/stu905}{{\em \mnras} {\bfseries 442}
  no.~3, (Aug., 2014) 2131--2144},
  \href{http://arxiv.org/abs/1311.7100}{{\ttfamily arXiv:1311.7100
  [astro-ph.CO]}}.

\bibitem{cooray/sheth}
A.~Cooray and R.~K. Sheth, ``{Halo models of large scale structure},''
  \href{http://dx.doi.org/10.1016/S0370-1573(02)00276-4}{{\em Phys.Rept.}
  {\bfseries 372} (2002) 1--129},
\href{http://arxiv.org/abs/astro-ph/0206508}{{\ttfamily arXiv:astro-ph/0206508
  [astro-ph]}}.

\bibitem{1997MNRAS.286..795K}
G.~{Kauffmann}, A.~{Nusser}, and M.~{Steinmetz}, ``{Galaxy formation and
  large-scale bias},'' \href{http://dx.doi.org/10.1093/mnras/286.4.795}{{\em
  \mnras} {\bfseries 286} no.~4, (Apr., 1997) 795--811},
  \href{http://arxiv.org/abs/astro-ph/9512009}{{\ttfamily
  arXiv:astro-ph/9512009 [astro-ph]}}.

\bibitem{2000MNRAS.318..203S}
U.~{Seljak}, ``{Analytic model for galaxy and dark matter clustering},''
  \href{http://dx.doi.org/10.1046/j.1365-8711.2000.03715.x}{{\em \mnras}
  {\bfseries 318} no.~1, (Oct., 2000) 203--213},
  \href{http://arxiv.org/abs/astro-ph/0001493}{{\ttfamily
  arXiv:astro-ph/0001493 [astro-ph]}}.

\bibitem{2000MNRAS.318.1144P}
J.~A. {Peacock} and R.~E. {Smith}, ``{Halo occupation numbers and galaxy
  bias},'' \href{http://dx.doi.org/10.1046/j.1365-8711.2000.03779.x}{{\em
  \mnras} {\bfseries 318} no.~4, (Nov., 2000) 1144--1156},
  \href{http://arxiv.org/abs/astro-ph/0005010}{{\ttfamily
  arXiv:astro-ph/0005010 [astro-ph]}}.

\bibitem{2003ApJ...593....1B}
A.~A. {Berlind}, D.~H. {Weinberg}, A.~J. {Benson}, C.~M. {Baugh}, S.~{Cole},
  R.~{Dav{\'e}}, C.~S. {Frenk}, A.~{Jenkins}, N.~{Katz}, and C.~G. {Lacey},
  ``{The Halo Occupation Distribution and the Physics of Galaxy Formation},''
  \href{http://dx.doi.org/10.1086/376517}{{\em \apj} {\bfseries 593} no.~1,
  (Aug., 2003) 1--25}, \href{http://arxiv.org/abs/astro-ph/0212357}{{\ttfamily
  arXiv:astro-ph/0212357 [astro-ph]}}.

\bibitem{2004ApJ...609...35K}
A.~V. {Kravtsov}, A.~A. {Berlind}, R.~H. {Wechsler}, A.~A. {Klypin},
  S.~{Gottl{\"o}ber}, B.~o. {Allgood}, and J.~R. {Primack}, ``{The Dark Side of
  the Halo Occupation Distribution},''
  \href{http://dx.doi.org/10.1086/420959}{{\em \apj} {\bfseries 609} no.~1,
  (July, 2004) 35--49}, \href{http://arxiv.org/abs/astro-ph/0308519}{{\ttfamily
  arXiv:astro-ph/0308519 [astro-ph]}}.

\bibitem{2020arXiv201204637V}
R.~{Voivodic} and A.~{Barreira}, ``{Responses of Halo Occupation Distributions:
  a new ingredient in the halo model \& the impact on galaxy bias},'' {\em
  arXiv e-prints} (Dec., 2020) arXiv:2012.04637,
  \href{http://arxiv.org/abs/2012.04637}{{\ttfamily arXiv:2012.04637
  [astro-ph.CO]}}.

\bibitem{2019MNRAS.488.2079B}
A.~{Barreira}, D.~{Nelson}, A.~{Pillepich}, V.~{Springel}, F.~{Schmidt},
  R.~{Pakmor}, L.~{Hernquist}, and M.~{Vogelsberger}, ``{Separate Universe
  simulations with IllustrisTNG: baryonic effects on power spectrum responses
  and higher-order statistics},''
  \href{http://dx.doi.org/10.1093/mnras/stz1807}{{\em \mnras} {\bfseries 488}
  no.~2, (Sept., 2019) 2079--2092},
  \href{http://arxiv.org/abs/1904.02070}{{\ttfamily arXiv:1904.02070
  [astro-ph.CO]}}.

\bibitem{2020JCAP...05..042I}
M.~M. {Ivanov}, M.~{Simonovi{\'c}}, and M.~{Zaldarriaga}, ``{Cosmological
  parameters from the BOSS galaxy power spectrum},''
  \href{http://dx.doi.org/10.1088/1475-7516/2020/05/042}{{\em \jcap} {\bfseries
  2020} no.~5, (May, 2020) 042},
  \href{http://arxiv.org/abs/1909.05277}{{\ttfamily arXiv:1909.05277
  [astro-ph.CO]}}.

\bibitem{2020A&A...633L..10T}
T.~{Tr{\"o}ster}, A.~G. {S{\'a}nchez}, M.~{Asgari}, C.~{Blake}, M.~{Crocce},
  C.~{Heymans}, H.~{Hildebrandt}, B.~{Joachimi}, S.~{Joudaki}, A.~{Kannawadi},
  C.-A. {Lin}, and A.~{Wright}, ``{Cosmology from large-scale structure.
  Constraining {\ensuremath{\Lambda}}CDM with BOSS},''
  \href{http://dx.doi.org/10.1051/0004-6361/201936772}{{\em \aap} {\bfseries
  633} (Jan., 2020) L10}, \href{http://arxiv.org/abs/1909.11006}{{\ttfamily
  arXiv:1909.11006 [astro-ph.CO]}}.

\bibitem{2021A&A...646A.140H}
C.~{Heymans}, T.~{Tr{\"o}ster}, M.~{Asgari}, C.~{Blake}, H.~{Hildebrandt},
  B.~{Joachimi}, K.~{Kuijken}, C.-A. {Lin}, A.~G. {S{\'a}nchez}, J.~L. {van den
  Busch}, A.~H. {Wright}, A.~{Amon}, M.~{Bilicki}, J.~{de Jong}, M.~{Crocce},
  A.~{Dvornik}, T.~{Erben}, M.~C. {Fortuna}, F.~{Getman}, B.~{Giblin},
  K.~{Glazebrook}, H.~{Hoekstra}, S.~{Joudaki}, A.~{Kannawadi},
  F.~{K{\"o}hlinger}, C.~{Lidman}, L.~{Miller}, N.~R. {Napolitano},
  D.~{Parkinson}, P.~{Schneider}, H.~{Shan}, E.~A. {Valentijn}, G.~{Verdoes
  Kleijn}, and C.~{Wolf}, ``{KiDS-1000 Cosmology: Multi-probe weak
  gravitational lensing and spectroscopic galaxy clustering constraints},''
  \href{http://dx.doi.org/10.1051/0004-6361/202039063}{{\em \aap} {\bfseries
  646} (Feb., 2021) A140}, \href{http://arxiv.org/abs/2007.15632}{{\ttfamily
  arXiv:2007.15632 [astro-ph.CO]}}.

\bibitem{2021arXiv210612580I}
M.~M. {Ivanov}, ``{Cosmological constraints from the power spectrum of eBOSS
  emission line galaxies},'' {\em arXiv e-prints} (June, 2021)
  arXiv:2106.12580, \href{http://arxiv.org/abs/2106.12580}{{\ttfamily
  arXiv:2106.12580 [astro-ph.CO]}}.

\bibitem{2021arXiv210513549D}
{DES Collaboration}, ``{Dark Energy Survey Year 3 Results: Cosmological
  Constraints from Galaxy Clustering and Weak Lensing},'' {\em arXiv e-prints}
  (May, 2021) arXiv:2105.13549,
  \href{http://arxiv.org/abs/2105.13549}{{\ttfamily arXiv:2105.13549
  [astro-ph.CO]}}.

\bibitem{2021arXiv210513548K}
E.~{Krause} and {et al}, ``{Dark Energy Survey Year 3 Results: Multi-Probe
  Modeling Strategy and Validation},'' {\em arXiv e-prints} (May, 2021)
  arXiv:2105.13548, \href{http://arxiv.org/abs/2105.13548}{{\ttfamily
  arXiv:2105.13548 [astro-ph.CO]}}.

\bibitem{baldauf/etal:2015}
T.~{Baldauf}, U.~{Seljak}, L.~{Senatore}, and M.~{Zaldarriaga}, ``{Linear
  response to long wavelength fluctuations using curvature simulations},''
  \href{http://dx.doi.org/10.1088/1475-7516/2016/09/007}{{\em \jcap} {\bfseries
  9} (Sept., 2016) 007}, \href{http://arxiv.org/abs/1511.01465}{{\ttfamily
  arXiv:1511.01465}}.

\bibitem{li/hu/takada:2016}
Y.~{Li}, W.~{Hu}, and M.~{Takada}, ``{Separate universe consistency relation
  and calibration of halo bias},''
  \href{http://dx.doi.org/10.1103/PhysRevD.93.063507}{{\em \prd} {\bfseries 93}
  no.~6, (Mar., 2016) 063507},
  \href{http://arxiv.org/abs/1511.01454}{{\ttfamily arXiv:1511.01454}}.

\bibitem{2018PhRvD..97l3526C}
C.-T. {Chiang}, W.~{Hu}, Y.~{Li}, and M.~{LoVerde}, ``{Scale-dependent bias and
  bispectrum in neutrino separate universe simulations},''
  \href{http://dx.doi.org/10.1103/PhysRevD.97.123526}{{\em \prd} {\bfseries 97}
  no.~12, (Jun, 2018) 123526},
  \href{http://arxiv.org/abs/1710.01310}{{\ttfamily arXiv:1710.01310
  [astro-ph.CO]}}.

\bibitem{andreas}
A.~S. {Schmidt}, S.~D.~M. {White}, F.~{Schmidt}, and J.~{St{\"u}cker},
  ``{Cosmological N-body simulations with a large-scale tidal field},''
  \href{http://dx.doi.org/10.1093/mnras/sty1430}{{\em \mnras} {\bfseries 479}
  (Sep, 2018) 162--170}, \href{http://arxiv.org/abs/1803.03274}{{\ttfamily
  arXiv:1803.03274 [astro-ph.CO]}}.

\bibitem{2020arXiv201106584A}
K.~{Akitsu}, Y.~{Li}, and T.~{Okumura}, ``{Cosmological simulation in tides:
  power spectrum and halo shape responses, and shape assembly bias},'' {\em
  arXiv e-prints} (Nov., 2020) arXiv:2011.06584,
  \href{http://arxiv.org/abs/2011.06584}{{\ttfamily arXiv:2011.06584
  [astro-ph.CO]}}.

\bibitem{2020MNRAS.496..483M}
S.~{Masaki}, T.~{Nishimichi}, and M.~{Takada}, ``{Anisotropic separate universe
  simulations},'' \href{http://dx.doi.org/10.1093/mnras/staa1579}{{\em \mnras}
  {\bfseries 496} no.~1, (July, 2020) 483--496},
  \href{http://arxiv.org/abs/2003.10052}{{\ttfamily arXiv:2003.10052
  [astro-ph.CO]}}.

\bibitem{2021MNRAS.503.1473S}
J.~{St{\"u}cker}, A.~S. {Schmidt}, S.~D.~M. {White}, F.~{Schmidt}, and
  O.~{Hahn}, ``{Measuring the tidal response of structure formation:
  anisotropic separate universe simulations using TREEPM},''
  \href{http://dx.doi.org/10.1093/mnras/stab473}{{\em \mnras} {\bfseries 503}
  no.~1, (May, 2021) 1473--1489},
  \href{http://arxiv.org/abs/2003.06427}{{\ttfamily arXiv:2003.06427
  [astro-ph.CO]}}.

\end{thebibliography}\endgroup

\end{document}